\documentclass[journal=jacsat,manuscript=article]{achemso}

\usepackage{chemformula} 
\usepackage[T1]{fontenc} 

\usepackage{graphicx}   

\usepackage{natbib} 

\usepackage[hidelinks]{hyperref}

\usepackage{xcolor}

\usepackage{amssymb}

\usepackage{caption}

\usepackage{soul}

\author{Yuhao Zhou}
\affiliation[Fudan University]
{State Key Laboratory of Molecular Engineering of Polymers, 
Research Center of AI for Polymer Science,
Key Laboratory of Computational Physical Sciences, 
Department of Macromolecular Science, 
Fudan University, Shanghai 200433, China}

\author{Huangyan Shen}
\affiliation[Fudan University]
{State Key Laboratory of Molecular Engineering of Polymers, 
Research Center of AI for Polymer Science,
Key Laboratory of Computational Physical Sciences, 
Department of Macromolecular Science, 
Fudan University, Shanghai 200433, China}

\author{Qingliang Song}
\affiliation[Fudan University]
{State Key Laboratory of Molecular Engineering of Polymers, 
Research Center of AI for Polymer Science,
Key Laboratory of Computational Physical Sciences, 
Department of Macromolecular Science, 
Fudan University, Shanghai 200433, China}

\author{Qingshu Dong}
\email{qsdong@fudan.edu.cn}
\affiliation[Fudan University]
{State Key Laboratory of Molecular Engineering of Polymers, 
Research Center of AI for Polymer Science,
Key Laboratory of Computational Physical Sciences, 
Department of Macromolecular Science, 
Fudan University, Shanghai 200433, China}

\author{Jianfeng Li}
\email{lijf@fudan.edu.cn}
\affiliation[Fudan University]
{State Key Laboratory of Molecular Engineering of Polymers, 
Research Center of AI for Polymer Science,
Key Laboratory of Computational Physical Sciences, 
Department of Macromolecular Science, 
Fudan University, Shanghai 200433, China}

\author{Weihua Li}
\email{weihuali@fudan.edu.cn}
\affiliation[Fudan University]
{State Key Laboratory of Molecular Engineering of Polymers, 
Research Center of AI for Polymer Science,
Key Laboratory of Computational Physical Sciences, 
Department of Macromolecular Science, 
Fudan University, Shanghai 200433, China}

\title{Fully automated inverse co-optimization of templates and block copolymer blending recipes for DSA lithography}

\abbreviations{DSA,BCP,BO}

\begin{document}

\begin{tocentry}

\centering
\includegraphics{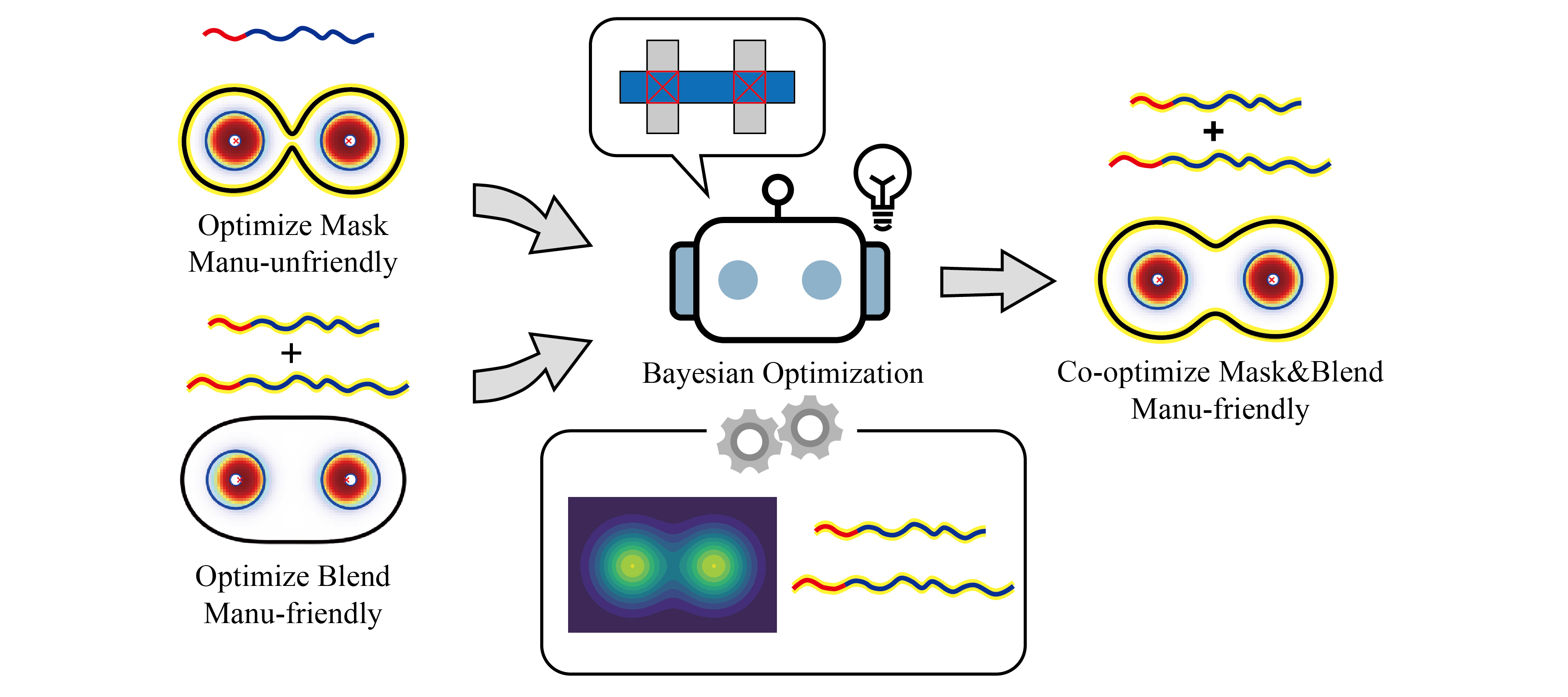}

\end{tocentry}

\begin{abstract}
    
The directed self-assembly (DSA) of block copolymers (BCPs) offers a highly promising approach for the fabrication of contact holes or vertical interconnect access at sub-7nm technology nodes. To fabricate circular holes with precisely controlled
size and positions, the self-assembly of block copolymers requires guidance from a
properly designed template. Effectively parameterizing the template shape to enable efficient optimization remains a critical yet challenging problem. Moreover, the optimized template must possess excellent manufacturability for practical applications.
In this work, we propose a Gaussian descriptor for characterizing the template
shape with only two parameters. We further propose to use AB/AB binary blends
instead of pure diblock copolymer to improve the adaptability of the block copolymer
system to the template shape. The Bayesian optimization (BO) is applied to
co-optimize the binary blend and the template shape. Our results demonstrate
that BO based on the Gaussian descriptor can efficiently yield the optimal
templates for diverse multi-hole patterns, all leading to highly matched
self-assembled morphologies. Moreover, by imposing constraints on the variation
of curvature of the template during optimization, superior manufacturability
is ensured for each optimized template. It is noteworthy that each key
parameter of the blend exhibits a relatively wide tunable window
under the requirement of rather high precision. Our work provides valuable
insights for advancing DSA technology, and thus potentially propels its practical applications forward.

\end{abstract}


With the continuous increase in demand for chip performance, the integration density of chips needs to be constantly improved, which has largely followed Moore's Law over the past few decades.\cite{RN30} The core technology that determines integration density is lithography.\cite{RN67, RN68} The most advanced lithography technique is Extreme Ultraviolet (EUV) with a wavelength of only 13.5 nanometers.\cite{RN70} Nevertheless, it remains essential to develop other advanced lithography techniques. On the one hand, EUV equipment is still very expensive,\cite{RN69} and its production capacity falls short of meeting the demand of semiconductor manufacturers,\cite{RN52} especially due to the explosive growth in computing power demands driven by AI. On the other hand, EUV lithography still lacks sufficient throughput when patterning high-precision features.\cite{RN50, RN51}

The directed self-assembly (DSA) of block copolymers is a bottom-up patterning technology that combines the spontaneous formation of ordered nanostructures by block copolymers with lithographic techniques.\cite{RN33, RN32, cao2015designing, gu2021epitaxial} The lithographic technique is applied to fabricate the guiding patterns that direct block copolymers to form the targeted patterns with precisely controlled size and shape at registered positions.\cite{RN35} In turn, the patterning resolution of DSA is primarily determined by the domain period of the self-assembled structures of block copolymer, which is essentially governed by its molecular weight. After decades of research, PS-$b$-PMMA diblock copolymer has emerged as the standard material for DSA, whose periodicity is down to 24 nanometers and is capable of meeting the precision requirements for sub-10-nanometer technology node.\cite{RN54,xiong2016directed} The resolution of DSA can be further enhanced through the development of high-$\chi$ block copolymer materials.\cite{RN53, mishra2022gallol, kwak2017fabrication} DSA has been successfully employed to fabricate various fundamental pattern elements in integrated circuits.\cite{RN15,RN5,RN6,RN7, RN71} Moreover, DSA has been applied to rectify the patterns made by EUV.\cite{RN47,RN48,RN49}

One of the most fundamental applications of DSA is to fabricate contact holes and vertical interconnect access (VIA) holes.\cite{RN35} A guiding template is first fabricated using lithographic techniques. Under the confinement of the guiding template, block copolymers self-assemble into vertically penetrated cylinders. Through selective etching, these cylinders are pattern-transferred to form nanoholes in the silicon substrate.\cite{RN36} Apparently, the size of nanoholes, which is considerably smaller than that of the template, is determined by the size of the cylinder. While the size of the cylinder is controlled by the parameters of the block copolymer, like the volume fraction of the A-block of AB diblock copolymer ($f$), the polymerization degree $N$ and the Flory-Huggins parameter $\chi$ that characterizes the immiscibility between the two different species.\cite{cong2016understanding}

When applying DSA for nanohole fabrication, cylinder-forming diblock copolymer, of which $f$ is about $0.2<f<0.3$ in the intermediate segregation, is preferred.\cite{RN55,RN14} For a single circular contact hole, the template is also circular. It is relatively easy to estimate the molecular parameters of the block copolymer and the radius of the circular template for the target nanohole with a given radius based on the bulk self-assembly behavior of the block copolymer. However, when the target pattern becomes multiple holes with a specific arrangement, the guiding template is no longer simply circular. For instance, in the case of a double-hole pattern, the template takes on a racetrack-like or peanut-like shape, depending on the distance between the two holes.\cite{RN9,RN8} If the template has an unsuitable shape or size, it can cause the self-assembled cylinders of the block copolymer to deviate from the desired circularity, position, and dimensions of the holes, and may even lead to the formation of extraneous domains.\cite{RN13, RN37} Therefore, for multiple-hole patterns, it is essential not only to select appropriate block copolymer parameters but also to optimize the dimensions and shape of the guiding template based on the block copolymer characteristics. The simultaneous optimization of both block copolymer parameters and template geometry poses significant challenges due to the vast parameter space involved.\cite{zhang2019customizing} In experiments, each validation of parameters involves a complete workflow encompassing template fabrication, block copolymer self-assembly (including spin coating, annealing, and characterization), pattern transfer and so on. Therefore, it is formidably expensive for experiments to proceed the optimization. Developing computational methods to optimize the parameters of block copolymer and template geometry will significantly enhance efficiency and reduce costs.\cite{RN38,RN39}

Computational optimization of DSA parameters involves two critical steps. One step is to calculate the self-assembled nanostructures formed by a given block copolymer under a specific template. The other step is to inversely adjust either the block copolymer parameters or the guiding template geometry based on the deviations between the self-assembled nanostructures and the target pattern.\cite{RN77} By iteratively cycling through these two steps, optimal block copolymer and the corresponding guiding template can be achieved, which is expected to generate the desired pattern. However, whether the resulting pattern meets the required precision specifications cannot be fully guaranteed. The primary reason is that when the self-assembled structures corresponding to the target pattern significantly deviate from the bulk phase behavior, the block copolymer becomes fundamentally incapable of forming such structures. On the other hand, even if the obtained hole patterns meet the requirements, another issue may arise: the guiding template geometry may be challenging to fabricate by patterning techniques. To address these dual challenges, we propose utilizing an AB/AB diblock copolymer blend to replace pure AB diblock copolymer melt. Previous works have demonstrated that blends enable broader-range modulation of self-assembly behavior.\cite{RN8,RN41} Specifically, long and short B-blocks can respectively fill far and near spaces between the cylinders and template boundaries through local segregation, thereby significantly enhancing adaptability to template geometries.\cite{RN14}
In turn, the template shape can be smoothed by tailoring the AB/AB blend, improving its manufacturability. At the same time, the precision of the resulting pattern can be improved.

In this work, we investigate the self-assembly of AB/AB binary blends laterally guided by the geometric template, aiming to achieve various multiple circular nanoholes with desired arrangements. SCFT as one of the most successful methods for studying the self-assembly of block copolymers is applied to calculate the self-assembled nanostructures under the geometrical confinement of template that is modeled by a masking scheme.\cite{RN24,RN22,RN25,RN26} To date, there are a few methods for optimizing template geometries, each presenting distinct advantages and limitations.\cite{RN9,RN40,RN11,RN12,RN56, RN75} We propose to assign a two-dimensional Gaussian function for each hole, with its maximum point located at the hole center. By superimposing all these Gaussian functions and using their contour lines to define the template profile, we can optimize the template geometry simply by adjusting the contour value of the superimposed Gaussian function($\nu$) and its characteristic parameter of standard variation($\tau$). In a word, there are only two variables to be optimized for the template profile, drastically reducing the parameters required for describing the template shape. Combined with independent parameters for specifying the blend, there are at most seven adjustable parameters in total. We apply Bayesian optimization (BO) to fully explore the parameter space for the optimal block copolymer parameters as well as the template geometry. The objective function for BO is defined based on the discrepancy between the self-assembled morphology of the blend and the target nanohole pattern, consisting of circularity deviation $\mathcal{L}_{\rm cir}$, 
position accuracy $\mathcal{L}_{\rm pos}$ and redundancy distribution $\mathcal{L}_{\rm rd}$. 
Our results demonstrate that the proposed optimization approach can efficiently achieve a variety of target multi-hole patterns with high precision.
Meanwhile, the blending system enables an optimal trade-off between template fabricability and pattern fidelity.

\section{Results and discussion}

\subsection{Optimize the template or the AB/AB blend}

\begin{figure}[htbp]
    \centering
    \includegraphics[width=0.7\textwidth]{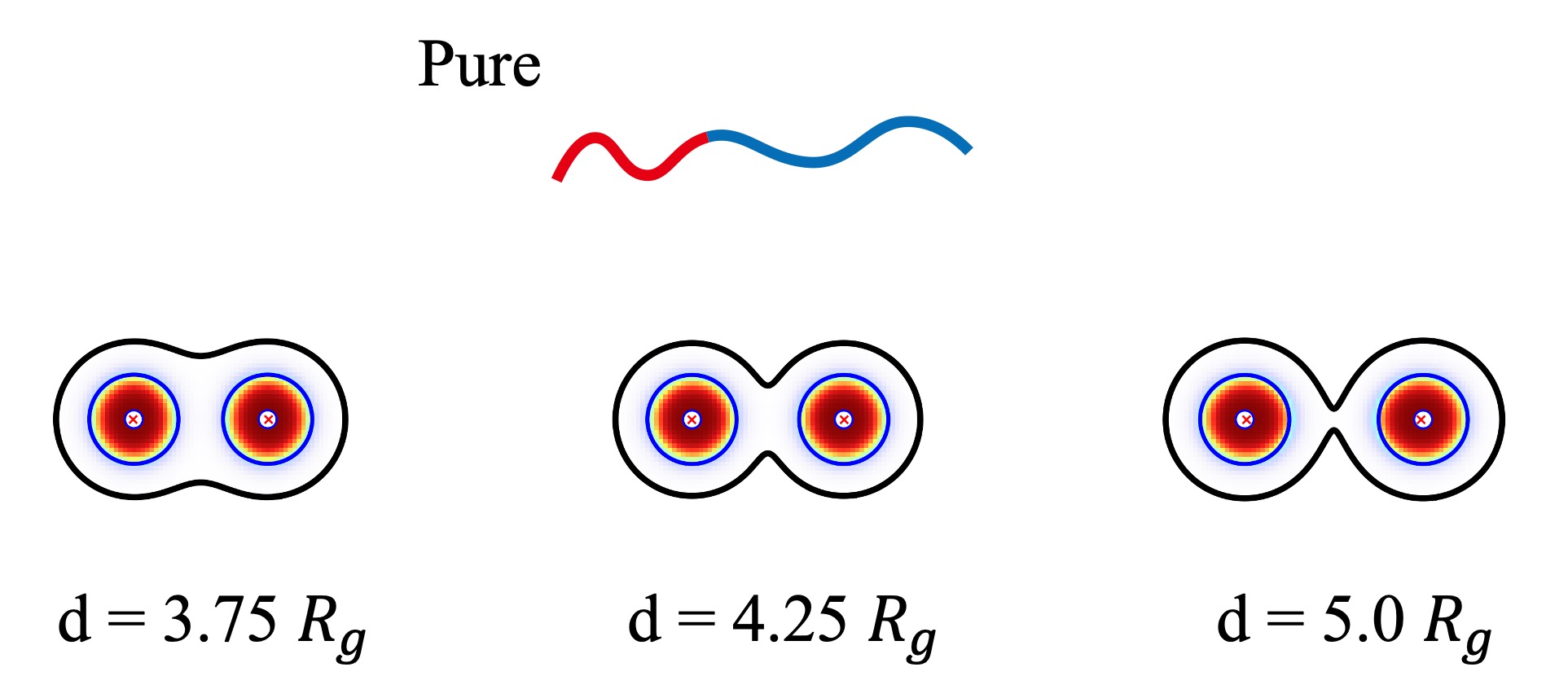}
    \caption{Template optimization with pure AB diblock: self-assembled morphologies
    inside the optimal template shapes. The black lines indicate the template boundaries, the blue outer circles represent the size of target holes, the blue inner circles indicate the target hole centers, and the red crosses mark the actual centers of the A-block domains (same notations apply below). }
    \label{mask_double}
\end{figure}

Double-hole is one of the fundamental units in the VIA layout, and therefore we take
it as an example to study the optimization process. When the separation of the two holes
is not large compared to their diameter, the template tends to be racetrack-like shape
for a pure cylinder-forming diblock copolymer melt. As the hole-to-hole distance
is increased, the racetrack-like shape gradually changes to a peanut-like shape.
When the waist of the peanut-like template becomes increasingly narrow, the concave part will grow sharper, reducing the manufacturability.
To demonstrate the change of the template shape with the separation of the two holes,
we consider three different distances, $d=3.75 R_g$, $4.25R_g$ and $5.00 R_g$, for a given diblock
copolymer with $\chi N=30$ and $f_{\rm A}=0.3$. The bulk cylinder-to-cylinder
distance and the diameter of the cylinder are $L_0=4.6 R_g$ and $d_0=2.6R_g$, respectively.
$d_0$ is estimated based on the point of maximum gradient in the volume-fraction
profile.

For each distance, we optimize the template shape, \textit{i.e.} the two
parameters $\tau$ and $\nu$ of the Gaussian descriptor, using BO to minimize objective function $\mathcal{L}_{\rm{total}}$
(see \textbf{\nameref{sec:method}} section).
Three optimized templates shapes for $d=3.75 R_g$, $4.25R_g$ and $5.00 R_g$ are shown
in Figure \ref{mask_double}.
For the three cases, the self-assembled morphologies of the block copolymers
within the optimized templates all match the target patterns perfectly,
including the circularity of the holes ($\mathcal{L}_{\rm{cir}}=1.4\times10^{-3}$, $1.1\times 10^{-3}$, $3.7\times 10^{-4}$) and positional accuracy ($\mathcal{L}_{\rm{pos}}=3.6\times 10^{-5}$, $7.1\times 10^{-6}$, $2.2\times10^{-4}$).
Apparently, no extraneous domains appear in the regions outside the holes,
resulting in low redundancy scores ($\mathcal{L}_{\rm{rd}}=1.1\times 10^{-4}$, $4.1\times10^{-4}$, $1.3\times 10^{-2}$).
Therefore, the value of multi-objective function is optimized to be
vanishingly low for each template ($<3\times10^{-3}$).

The template for $d=3.75R_g$ is close to a racetrack, while
that for $d=4.25R_g$ is more like peanut.
In contrast, the waist part of the template
for $d=5.00R_g$ becomes very narrow and sharp (\textit{i.e.} large variation of curvature),
which presents a big challenge
for manufacturing. The main reason is that when the two holes are widely separated,
the space near the intersection between the midplane of the two holes and the template
is too far for the B-block to fill if the template maintains the racetrack-like shape.
Consequently, the template autonomously evolves into a peanut-like shape during optimization, eliminating the region far away from the holes.
In other words, the neat block copolymer fails to accommodate the highly
non-uniform matrix thickness within the racetrack-like template.

\begin{figure}[htbp]
    \centering
    \includegraphics[width=1\textwidth]{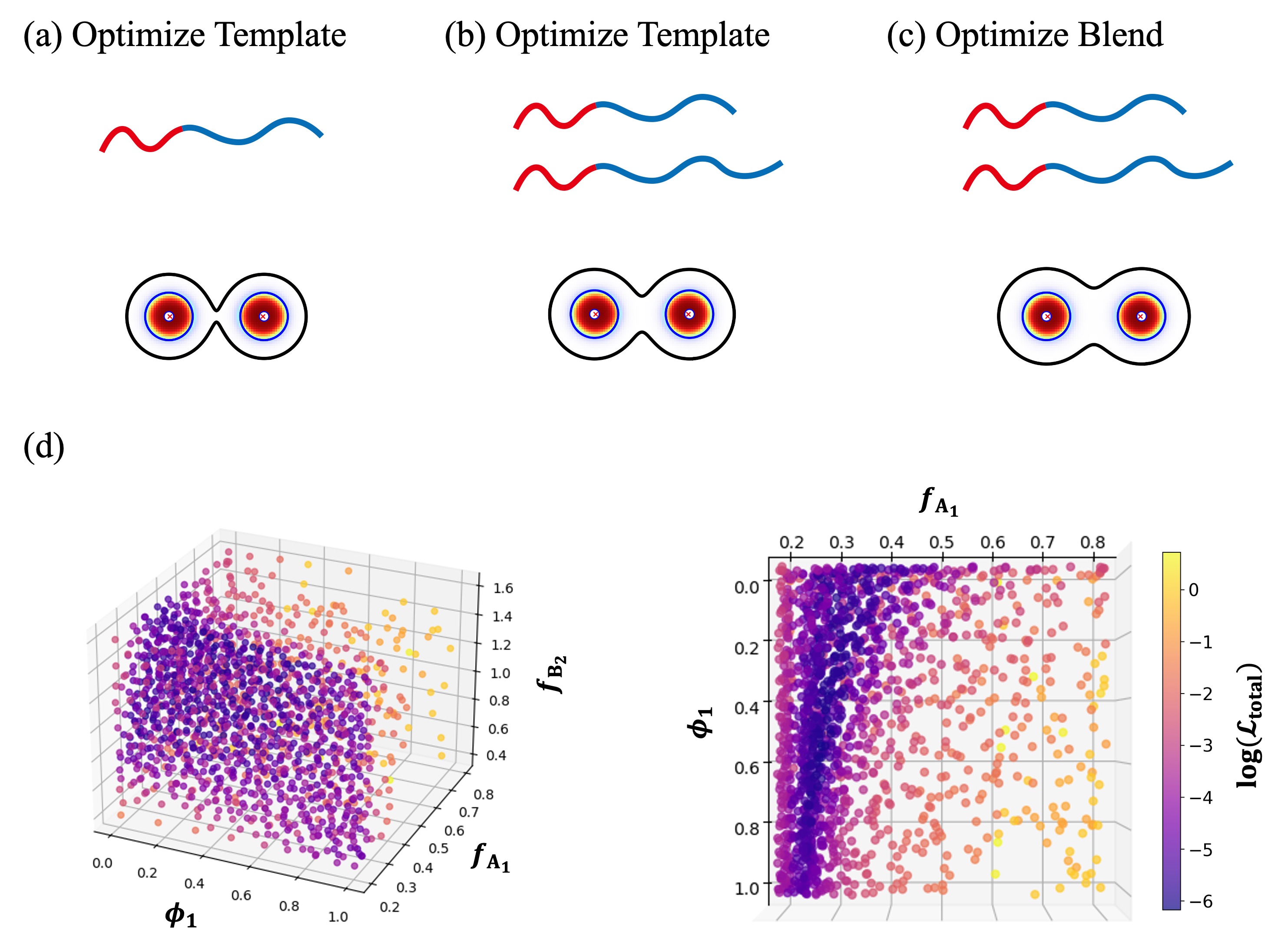}
    \caption{Optimize the template or the blend for targeting the double-hole pattern
    with $d=5.00R_g$. (a) the self-assembled morphology of pure AB diblock
    copolymer in the optimized template. (b) the self-assembled morphology of 
    the AB/AB blend in the optimized template. (c) the self-assembled morphology of the optimized AB/AB blend in a specific template. (d) visualization of the multi-objective function in the 3D parameter space of the AB/AB blend: The color represents the logarithm of the objective function value, with darker color indicating lower (better) objective function value.}
    \label{blend_double_fit}
\end{figure}

In principle, adding another diblock copolymer with longer B-block into the template would enable the longer B-block to fill the regions far away from the holes, thereby
enhancing the adaptability of the block copolymer system to the template shape.
To verify this speculation, we use an AB/AB binary blend with same A-block length but different B-block length to replace the pure diblock copolymer. One diblock copolymer is set as reference, and its total number of segments on each chain is $N$ ($\chi N=30$), 
composed of $N_{\rm A_1}=f_{\rm A_1}N$ A-segments ($f_{\rm A_1}=0.3$). The number of A-segments on the other diblock copolymer is chosen to be the same as that of the reference copolymer, $N_{\rm A_2}=f_{\rm A_1}N$, while the number of B-segments is larger, denoted by $N_{\rm B_2}=f_{\rm B_2}N$($f_{\rm B_2}=1.2$). The volume fraction of the reference diblock copolymer is set to $\phi_1=0.6$.
Figures \ref{blend_double_fit}a and b compare the optimized template
for the pure AB diblock copolymer and the AB/AB blend to target the
double-hole pattern with $d=5.00R_g$, respectively.
Apparently, the curvature variation of the template for the blend
is considerably reduced in comparison with that for the pure system,
while high precision is retained ($\mathcal{L}_{\rm pos}=2.9\times 10^{-4}$, $\mathcal{L}_{\rm cir}=6.7\times 10^{-4}$, $\mathcal{L}_{\rm rd}=1.1\times 10^{-2}$).
This comparison confirms that the AB/AB blend has greater adaptability to
the template shape. Alternatively, we can also optimize the blend system to
adapt to a specific template with reasonably low curvature variation. 
We generate a template via the Gaussian descriptor with $\tau=0.18$ and $\nu=0.12$, 
which has much thicker waist or much smaller variation of curvature 
and thus better manufacturability than the optimal template for the pure diblock copolymer to target the double-hole pattern with $d=5.00R_g$. 
With fixed $\chi N=30$ and $N_{\rm A_1}=N_{\rm A_2}$, BO is applied to 
optimize the remaining three variables, $f_{\rm A_1}$, $f_{\rm B_2}$, and $\phi_1$ of the AB/AB blend, to form the self-assembled morphology matching
the target double-hole pattern.

\begin{figure}[htbp]
    \centering
    \includegraphics[width=0.4\textwidth]{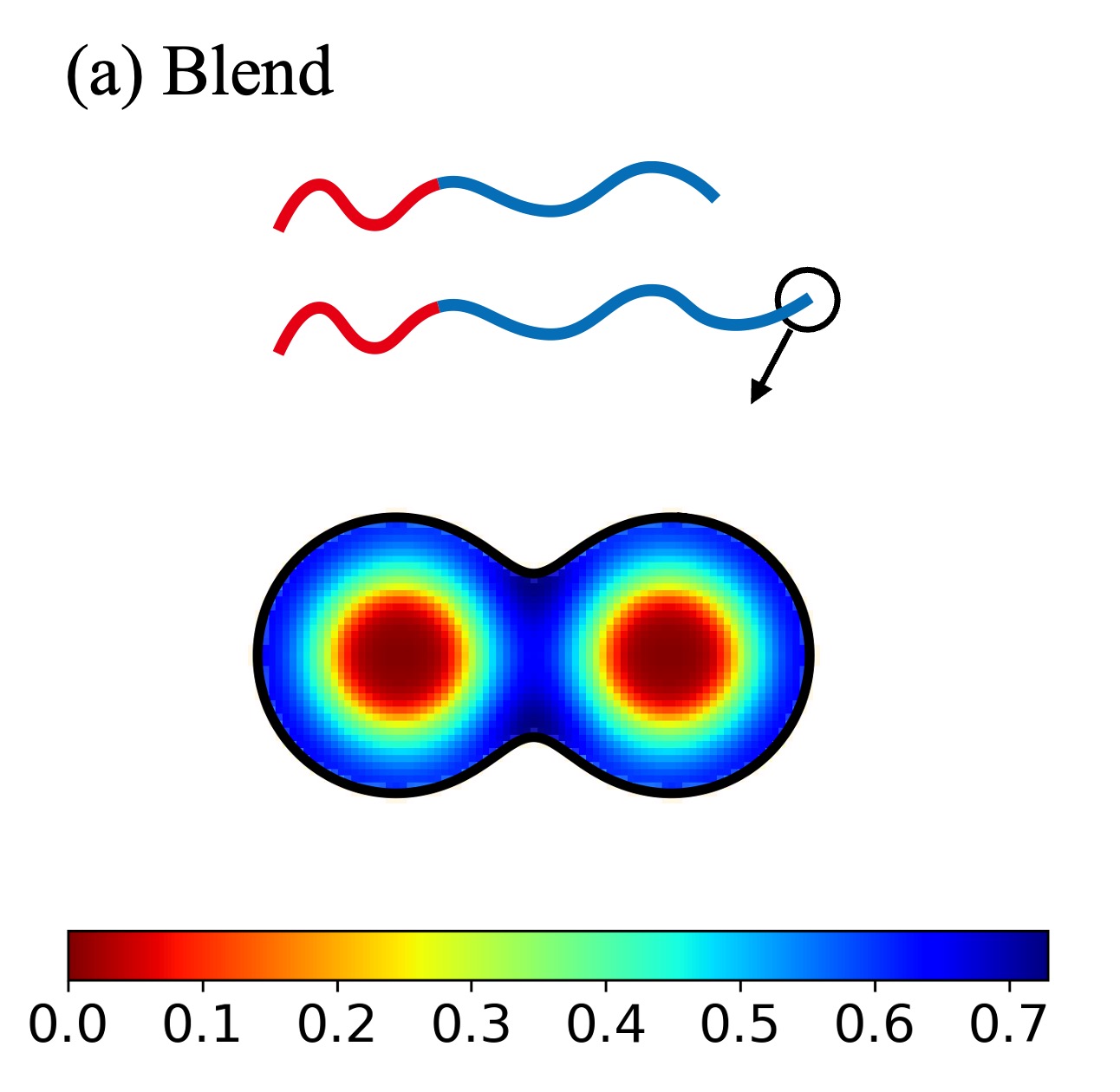}
    \caption{Spatial distribution of the free end of the long B-block from the second diblock copolymer in the binary blend.}
    \label{blend_double_fit_joint}
\end{figure}

The BO process is visualized by the scattering plot of the multi-objective
function in the explored parameter space (Figure \ref{blend_double_fit}), which exhibits
a rather wide processing window for the target pattern indicated by
dark-color spots. The optimal parameters suggested by BO
are $f_{\rm A_1}=0.278$, $f_{\rm B_2}=1.42$, and $\phi_1=0.55$,
and the resulting morphology matches the target
double-hole pattern very well ($\mathcal{L}_\mathrm{cir}=5.5\times10^{-4}$, $\mathcal{L}_\mathrm{pos}=2.9\times10^{-4}$, and $\mathcal{L}_\mathrm{rd}=8.6\times 10^{-3}$).
The processing window indicates that a longer B-block is required for
the second copolymer to effectively fill the far space within the template,
which is evidenced by the distribution of the B-end of the second copolymer
in Figure \ref{blend_double_fit_joint}.
Although the BO optimization results indicate that simultaneously optimizing the A and B blocks (or chain length and volume fraction) of pure diblock copolymer can also yield the self-assembled morphology that matches the target pattern reasonably, the blending system demonstrates superior performance.

\begin{figure}[htbp]
    \centering
    \includegraphics[width=1\textwidth]{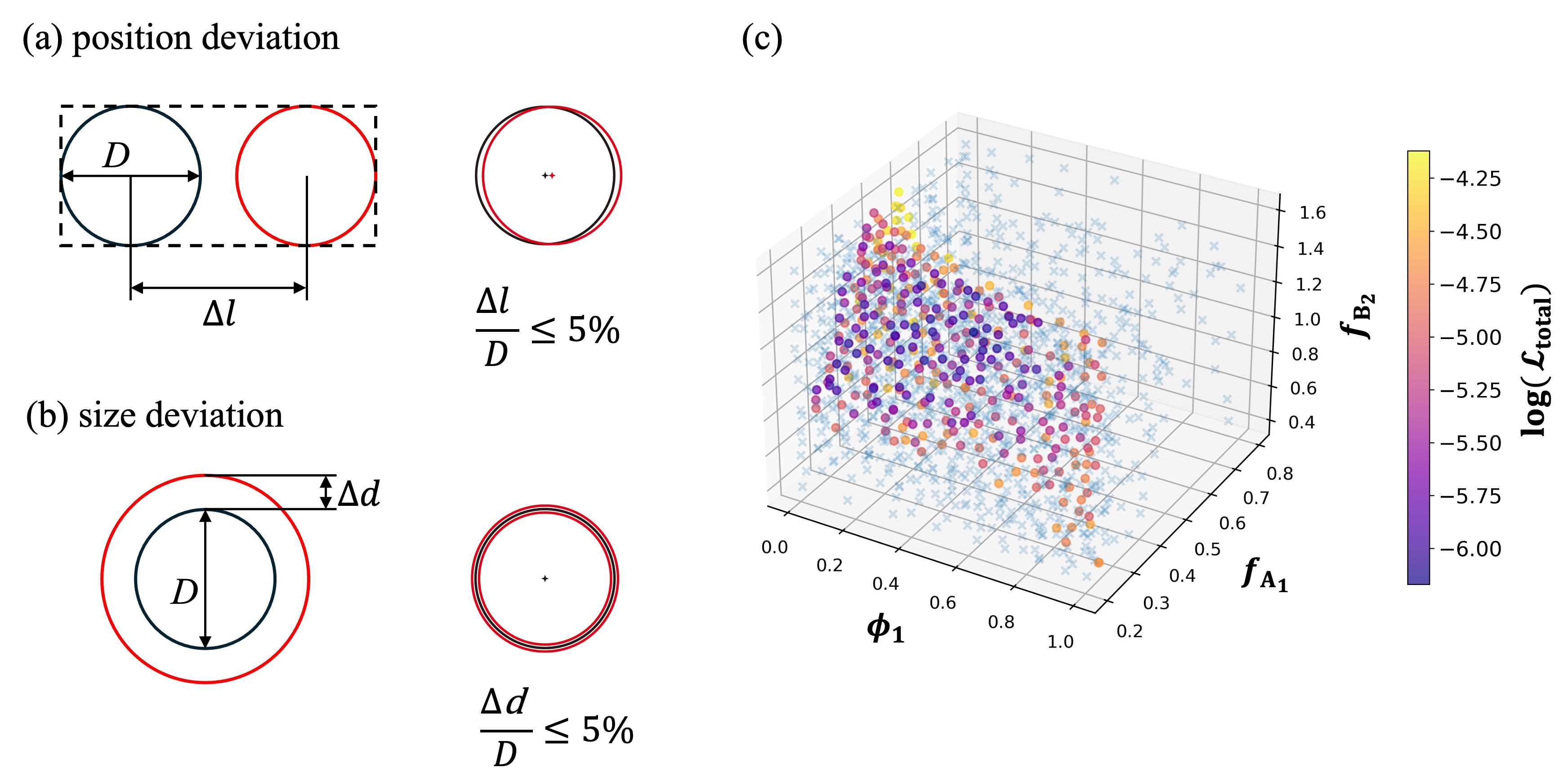}
    \caption{(a) position deviation in an ideal case (b) size deviation in an ideal case. (c) parameter window 
    that satisfies the conditions $\mathcal{L}_{\rm pos} \le 1.2\times 10^{-3}$ and $\mathcal{L}_{\rm cir} \le 2.5\times 10^{-3}$, where circular dots represent feasible points whose $\mathcal{L}_{\rm total}$ is indicated by color spectrum, whereas blue crosses indicate non-compliant points.}
    \label{score2ratio}
\end{figure}

In the following, we will elucidate how the the three metrics ($\mathcal{L}_{\rm pos}$, $\mathcal{L}_{\rm cir}$, $\mathcal{L}_{\rm rd}$) correspond to actual precisions. Since the third metric of the
redundancy distribution often decreases automatically in tandem with the other two,
we mainly discuss $\mathcal{L}_{\rm pos}$ and $\mathcal{L}_{\rm cir}$.
Two ideal cases are considered to see how $\mathcal{L}_{\rm pos}$ and $\mathcal{L}_{\rm cir}$ 
change with the deviations of position and size, respectively
as shown in Figure \ref{score2ratio}a and b.
One ideal case is that the hole is only shifted by $\Delta l$ from the target hole,
giving rise to $\mathcal{L}_{\rm{pos}}=\frac{(\Delta l/D)^2}{(\Delta l/D)^2+2\Delta l/D + 2}$ from eq \ref{eq:loss_pos}.
\begin{eqnarray}
\mathcal{L}_{\rm{pos}} &=& \frac{1}{n_{\rm hole}}\sum_{i=1}^{n_{\rm hole}}\frac{(x_{i}-\widehat{x_{i}})^2+(y_{i}-\widehat{y_{i}})^2}{(x_{i, \rm{left}}-x_{i, \rm{right}})^2+(y_{i, \rm{up}}-y_{i, \rm{down}})^2} \nonumber \\
&=& \frac{\Delta l^2}{(\Delta l + D)^2 + D^2} \nonumber \\
&=& \frac{(\frac{\Delta l}{D})^2}{(\frac{\Delta l}{D})^2+2\frac{\Delta l}{D} + 2}\label{eq:loss_pos}
\end{eqnarray}
where $D$ is the diameter of the target hole.
If $\Delta l/D=0.05$, corresponding to $\mathcal{L}_{\rm pos}=1.2\times 10^{-3}$, the positional error is only $\Delta l=1 {\rm nm}$
for $D=20 {\rm nm}$.
In the other case, it is assumed that the hole is ideally circular and
only has size deviation of $\Delta d$,
so that
\begin{eqnarray}
\mathcal{L}_{\rm{cir}}&=&\frac{1}{n_{\rm hole}}\sum_{i=1}^{n_{\rm hole}}\frac{(r_{i}-a_{i})^2+(r_{i}-b_{i})^2}{2r_{i}^2} \nonumber\\
&=&\frac{(\Delta d)^2+(\Delta d)^2}{2D^2} \nonumber\\
&=& \frac{(\Delta d)^2}{D^2} \label{eq:loss_cir}
\end{eqnarray}

\begin{figure}[htbp]
    \centering
    \includegraphics[width=1\textwidth]{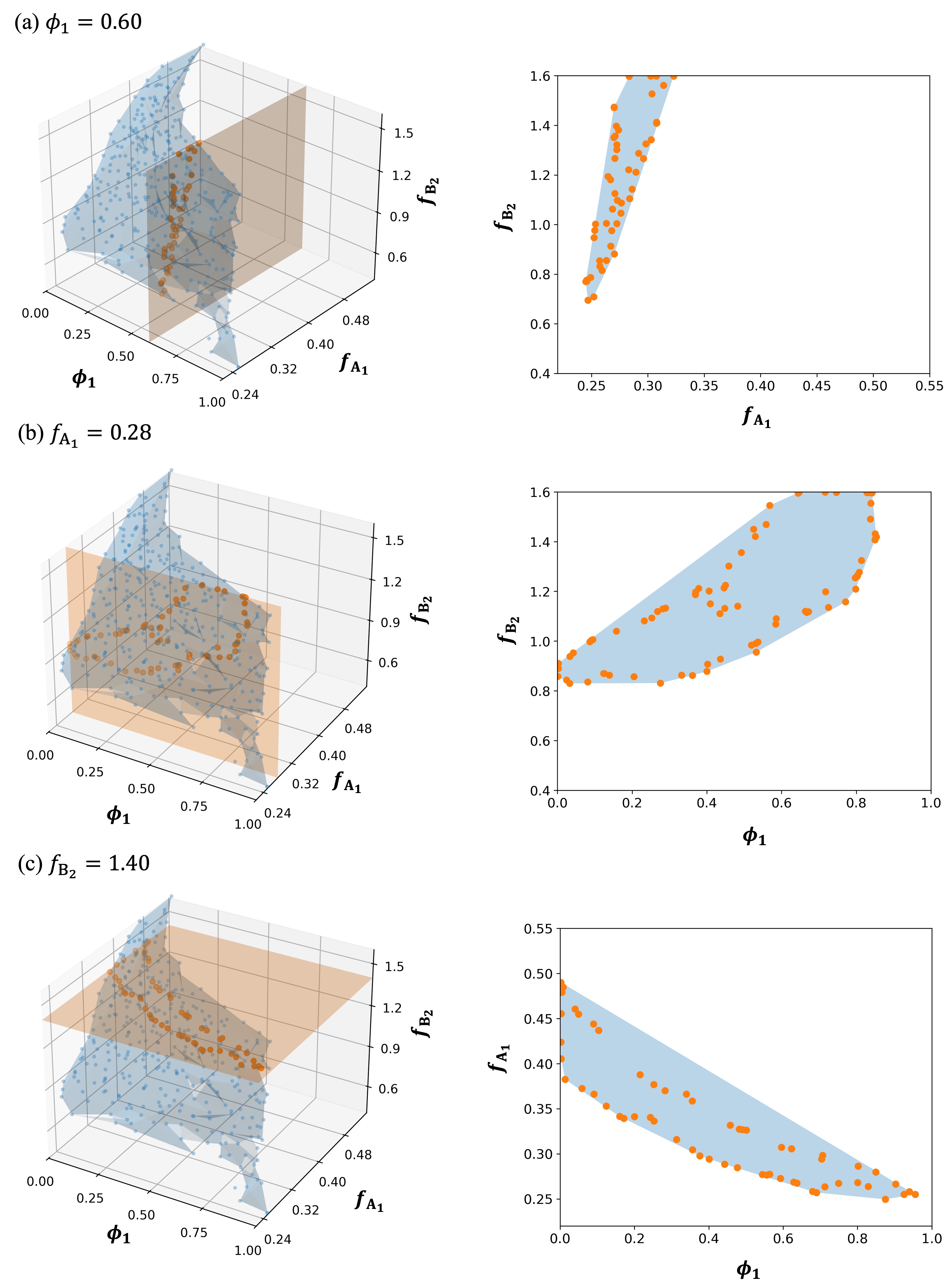}
    \caption{Representative two-dimensional parameter windows: (a) $f_{\rm A_1}$-$f_{\rm B_2}$ parameter window at $\phi_1=0.60$. (b) $\phi_1$-$f_{\rm B_2}$ parameter window at $f_{\rm A_1}=0.28$. (c) $\phi_1$-$f_{\rm A_1}$ parameter window at $f_{\rm B_2}=1.40$.}
    \label{process_window}
\end{figure}

For $\Delta d/D=0.05$, $\mathcal{L}_{\rm cir}=2.5\times 10^{-3}$.
If we take $\mathcal{L}_{\rm pos}=1.2\times 10^{-3}$ and $\mathcal{L}_{\rm cir}=2.5\times 10^{-3}$ in the two ideal cases as thresholds, we can
roughly estimate the parameter region of the binary blend
for the target double-hole pattern using a fixed peanut-like template,
which is indicated by the color spots in the
scattering plot of Figure \ref{score2ratio}c. In the main area of this parameter window,
the three parameters have rather broad adjustable ranges.
For example, as shown in Figure \ref{process_window}, when $\phi_1=0.60$ and $f_{\rm A_1}=0.28$, the parameter window of $f_{\rm B_2}$ is approximately $[1.1,1.5]$. For $\phi_1=0.60$ and $f_{\rm B_2}=1.40$, the parameter window of $f_{\rm A_1}$ is roughly $[0.27, 0.31]$. Finally, the parameter window of $\phi_1$ is as broad as
$0.50\lesssim \phi_1\lesssim 0.85$ for $f_{\rm A_1}=0.28$ and $f_{\rm B_2}=1.40$.

\begin{figure}[htbp]
    \centering
    \includegraphics[width=0.7\textwidth]{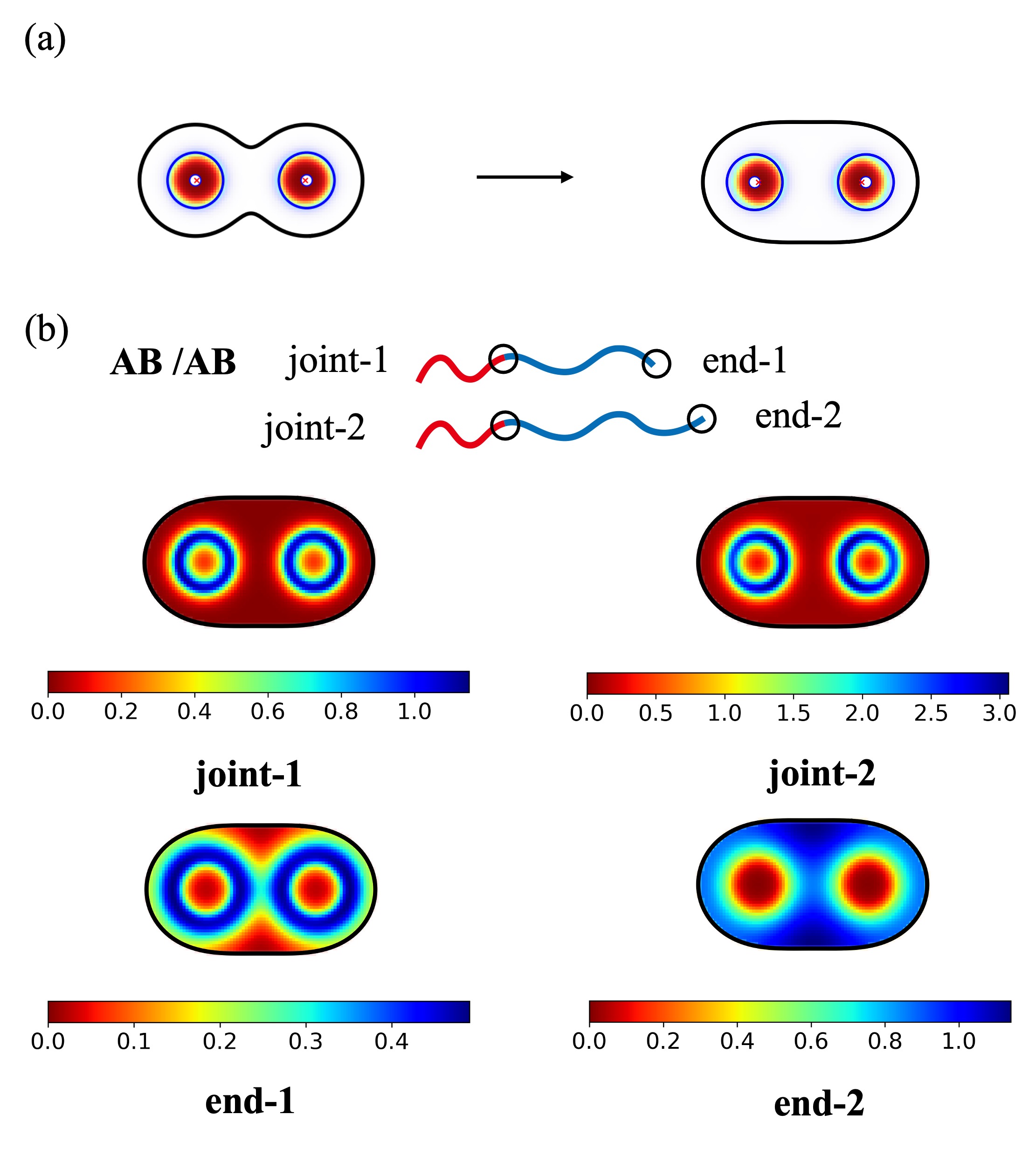}
    \caption{(a) Self-assembled morphologies of
    the optimized AB/AB blend in the
    specific template: (left) peanut-like, (right) racetrack-like. (b) spatial distributions of the AB junction points and the free ends of the B-blocks for both chains in the blend.}
    \label{blend_double_stadium_fit}
\end{figure}

The binary blend exhibits a better adaptability to templates than the
pure block copolymer, yet it cannot accommodate all templates, for example,
a racetrack-like template.
Figure \ref{blend_double_stadium_fit} compares the optimized results between the peanut-like template
and the racetrack-like template that is also generated by the Gaussian
descriptor with $\tau=0.65$ and $\nu=0.65$.
The optimal value of the multi-objective function for the racetrack-like template
is $\mathcal{L}_{\rm total}\approx 8.1 \times 10^{-3}$, which is remarkably higher than that of the peanut-like template ($\mathcal{L}_{\rm total}\approx 2.1\times 10^{-3}$).
Specifically, the self-assembled cylinders are notably smaller than the
target holes ($\mathcal{L}_{\rm cir}=6.2\times 10^{-3}>2.5\times 10^{-3}$), and their centers deviate from the hole centers significantly ($\mathcal{L}_{\rm pos}=2.5\times 10^{-3}>1.2\times 10^{-3}$).
The distributions of the junction points and the B-ends all confirm
that the long B-blocks undergo local segregation from the short B-blocks,
preferentially filling the far space at the mid-vertical plane in
the racetrack-like template.  
Therefore, it is necessary to simultaneously 
optimize the template shape and the binary blend.

\subsection{Co-optimization of the template and the AB/AB blend}

\begin{figure}[htbp]
    \centering
    \includegraphics[width=1.0\textwidth]{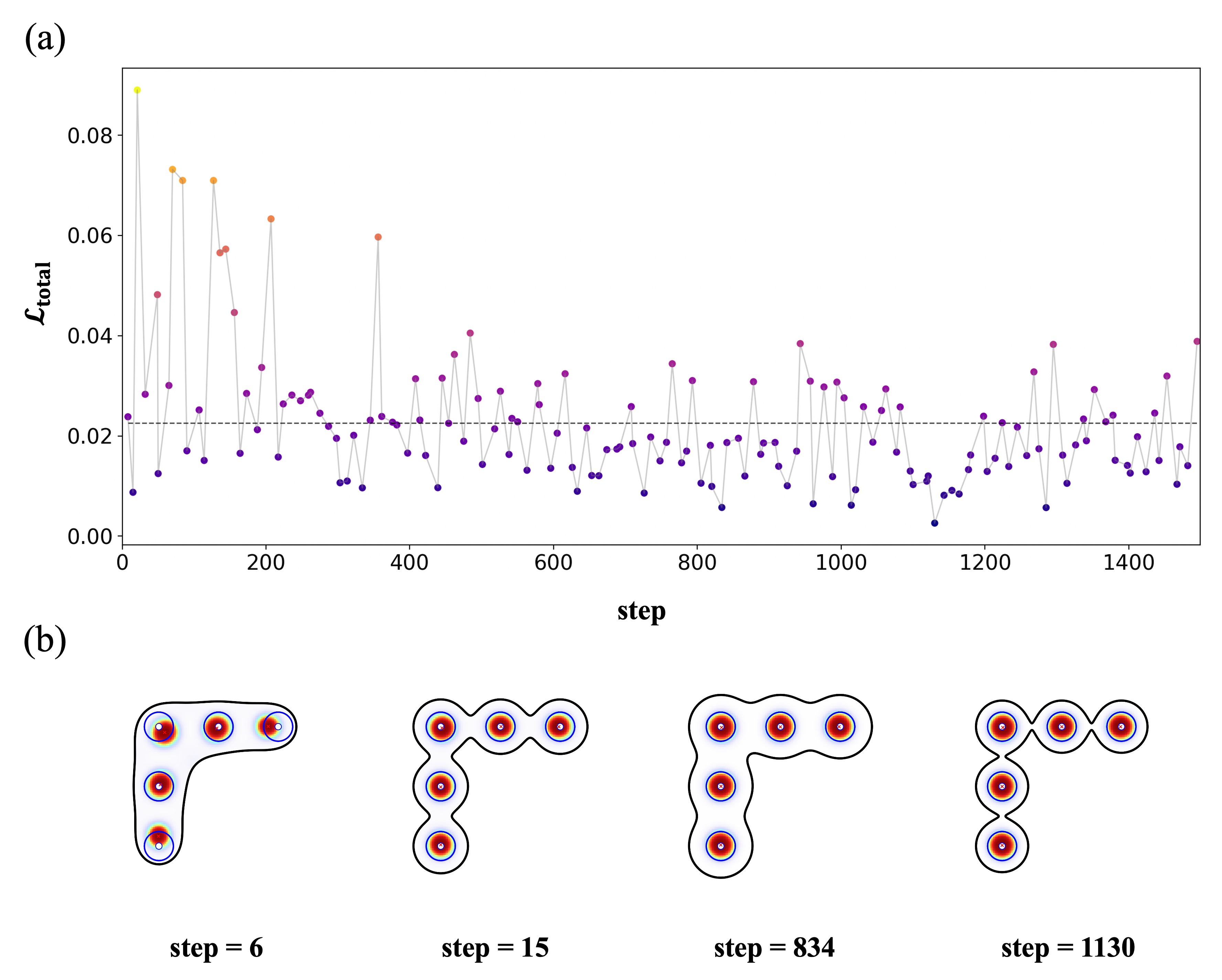}
    \caption{(a) Variation of the multi-objective function value during the co-optimization process for the L-shaped multi-hole target. For visualization clarity, the lowest value within every 10 iterations is plotted. (b) Self-assembled morphologies at typical steps.}
    \label{co_illustrate}
\end{figure}

\begin{table}[htbp]
  \centering
  \caption{Values of individual objective functions and parameters at iteration steps of 6, 15, 834 and 1130}
  \label{tab:obj_func}
  \begin{tabular}{|c|c|c|c|c|}
    \hline
    step & 6 & 15 & 834 & 1130\\
    \hline
    $\mathcal{L}_{\rm pos}$ & $2.44\times 10^{-2}$ & $8.75\times 10^{-3}$ & $4.92\times 10^{-4}$ & $2.98\times 10^{-4}$\\
    $\mathcal{L}_{\rm cir}$ & $1.33\times 10^{-2}$ & $8.78\times 10^{-4}$ & $5.25\times 10^{-4}$ & $1.13\times 10^{-3}$\\
    $\mathcal{L}_{\rm rd}$ & $2.07\times 10^{-1}$ & $8.09\times 10^{-3}$ & $1.79\times 10^{-2}$ & $6.86\times 10^{-3}$\\
    $\mathcal{L}_{\rm total}$ & $7.46\times 10^{-2}$ & $1.75\times 10^{-2}$ & $5.73\times 10^{-3}$ & $2.60\times 10^{-3}$\\
    $\phi_1$ & $0.256$ & $0.670$ & $0.006$ & $0.062$\\
    $f_{\rm A_1}$ & $0.134$ & $0.180$ & $0.185$ & $0.375$\\
    $f_{\rm A_2}$ & $0.269$ & $0.563$ & $0.269$ & $0.304$\\
    $f_{\rm B_2}$ & $0.789$ & $1.050$ & $1.109$ & $0.842$\\
    \hline
  \end{tabular}
\end{table}

To demonstrate the co-optimization effectiveness of the template and the blend,
we choose a rather complex target multi-hole pattern, composed of five holes
that form an L-arrangement. The diameter of each hole
and their neighboring distance are set as $2.25R_g$ and $4.625R_g$, respectively.
Differing from the double-hole pattern with two
mirror-symmetric holes, the five-hole layout belonging to the $D_1$ symmetry group
consists of three holes with non-equivalent positions,
raising the difficulty in the optimization,
especially about the template shape.
Therefore, the quintuple-hole pattern serves as a rigorous test case for evaluating the effectiveness of our co-optimization strategy.

Figure \ref{co_illustrate}a shows the variation of the multi-objective function with Bayesian optimization iterations, while Figure \ref{co_illustrate}b presents the self-assembled morphologies at
a few typical Bayesian optimization steps of $\rm{step}=6$, $15$, $834$, and $1130$.
At $\rm{step}=6$, the template shape can result in the formation of
five cylindrical domains, but the locations as well as the sizes of these cylinders
remarkably deviate from those of the target holes, especially those 
at the corner and at the ends. After a few more iterations like $\rm{step}=15$,
the matching degree between the self-assembled cylinders and the target holes
has been considerably improved. The template shape has been notably rectified. 
In fact, the parameters of the binary blend are optimized synchronously, that is,
$(f_{\rm A_1},f_{\rm A_2}, f_{\rm B_2}, \phi_1)=(0.134, 0.269, 0.789, 0.256)$ at $\rm step=6$ evolve to $(f_{\rm A_1},f_{\rm A_2}, f_{\rm B_2}, \phi_1)=(0.180, 0.563, 1.050, 0.670)$ at $\rm step=15$ as shown in Table \ref{tab:obj_func}.

With further optimization, the matching precision between the self-assembled
morphology and the target pattern is progressively improved. At ${\rm step=834}$,
both $\mathcal{L}_{\rm{pos}}$ and $\mathcal{L}_{\rm{cir}}$ are reduced to the order of $10^{-4}$, leading to
a reasonably high precision with $\mathcal{L}_{\rm{total}}=5.7\times 10^{-3}$.
During subsequent optimization steps, although the precision can be further improved, the optimization may compromise the manufacturability of the
template (\textit{e.g.} ${\rm step=1130}$).
\begin{figure}[htbp]
    \centering
    \includegraphics[width=1\textwidth]{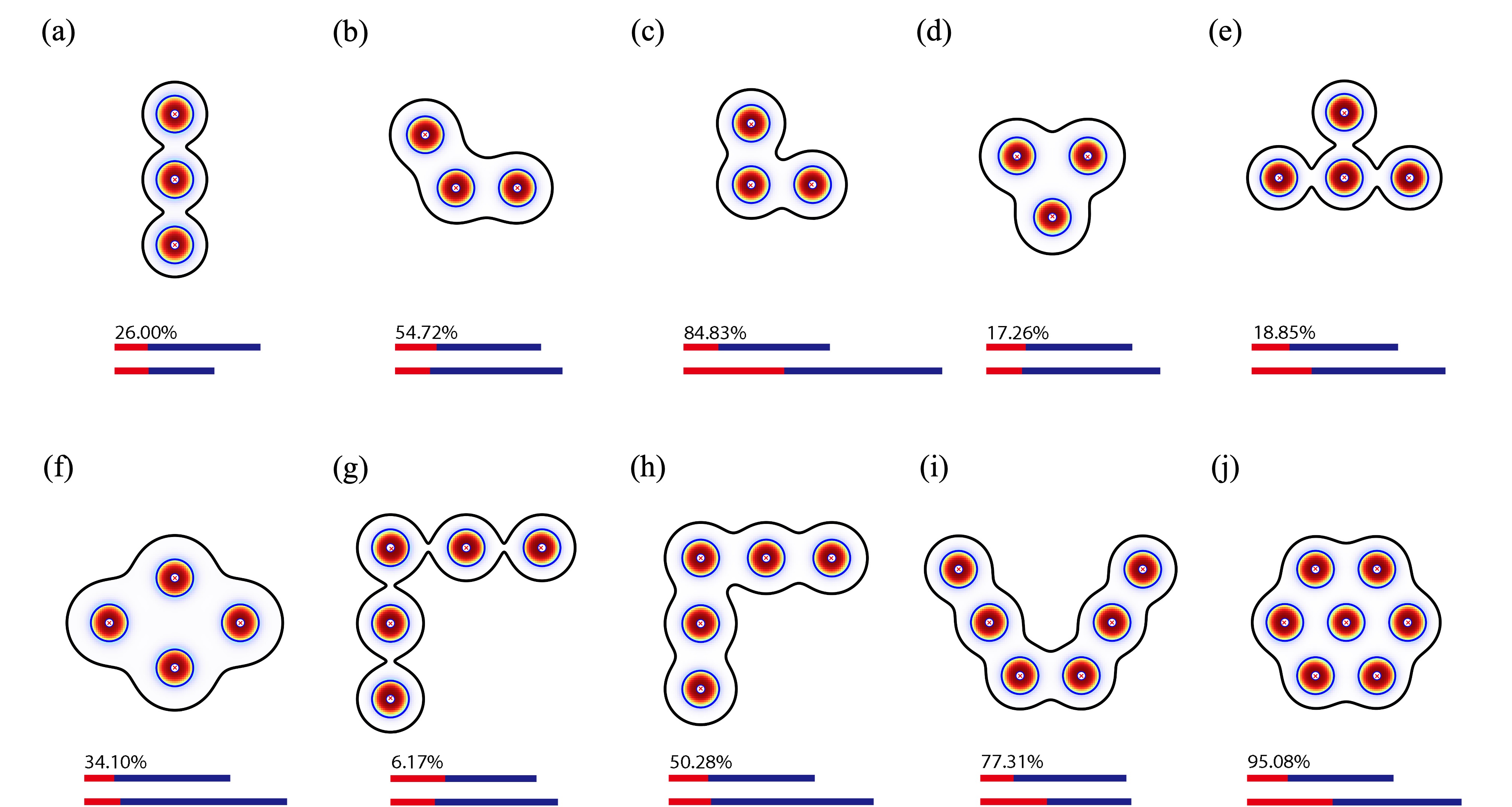}
    \caption{Self-assembled morphologies of the optimized AB/AB blend in the optimized template (Percentage means the volume fraction of first chain. Red line and blue line indicate A block and B block, respectively). (a-d) triple-hole, (e-f) quadruple-hole, (g–h) quintuple-hole, (i) sextuple-hole, and (j) septuple-hole.}
    \label{co_opt}
\end{figure}
We further consider a series of
distinct multi-hole patterns, including four triple-hole patterns,
two quadruple-hole patterns, two quintuple-hole patterns, one sextuple-hole 
pattern and one septuple-hole pattern. 
Figure \ref{co_opt} shows the optimization process 
can efficiently find the high-precision solution for each pattern. 
However, similar to previous quintuple-hole case (Figure \ref{co_opt}g), 
in Figures \ref{co_opt}a, e, h and i, the optimization sacrifices some manufacturability of the template in pursuit of higher precision.
As previously discussed, due to the limited manufacturing precision, 
the optimal solution corresponding to the highest accuracy is usually not
essential. Therefore, our optimization process should not merely pursue the optimal solution, but must also take the manufacturability of the template into account.

Since the manufacturability of the template is closely related to
the variation of curvature, we introduce a constraint to the latter
in our co-optimization approach. In the Gaussian descriptor, the
variation of curvature of the template can be simply constrained
by setting the threshold of $\tau$, like $\tau\ge 0.35$ that
generates the template shapes with reasonably small variation of curvature.
In the previous optimization process, one of the diblock copolymers
of the binary blend is chosen as the reference chain with fixed
number of segments $N$ (\textit{i.e.} $\chi N=30$). Since the length unit
$R_g=N^{1/2}b/\sqrt{6}$ is associated with $N$, it lowers the
adaptability of the binary blend to the absolute size of
the multi-hole patterns to some extent.
Therefore, we simultaneously adjust the numbers of segments ($N_1$ and $N_2$)
for both block copolymers, while setting $N$ as an independent reference parameter
with $\chi N=30$ and maintaining the length unit as $R_g=N^{1/2}b/\sqrt{6}$.

\begin{figure}[htbp]
    \centering
    \includegraphics[width=1\textwidth]{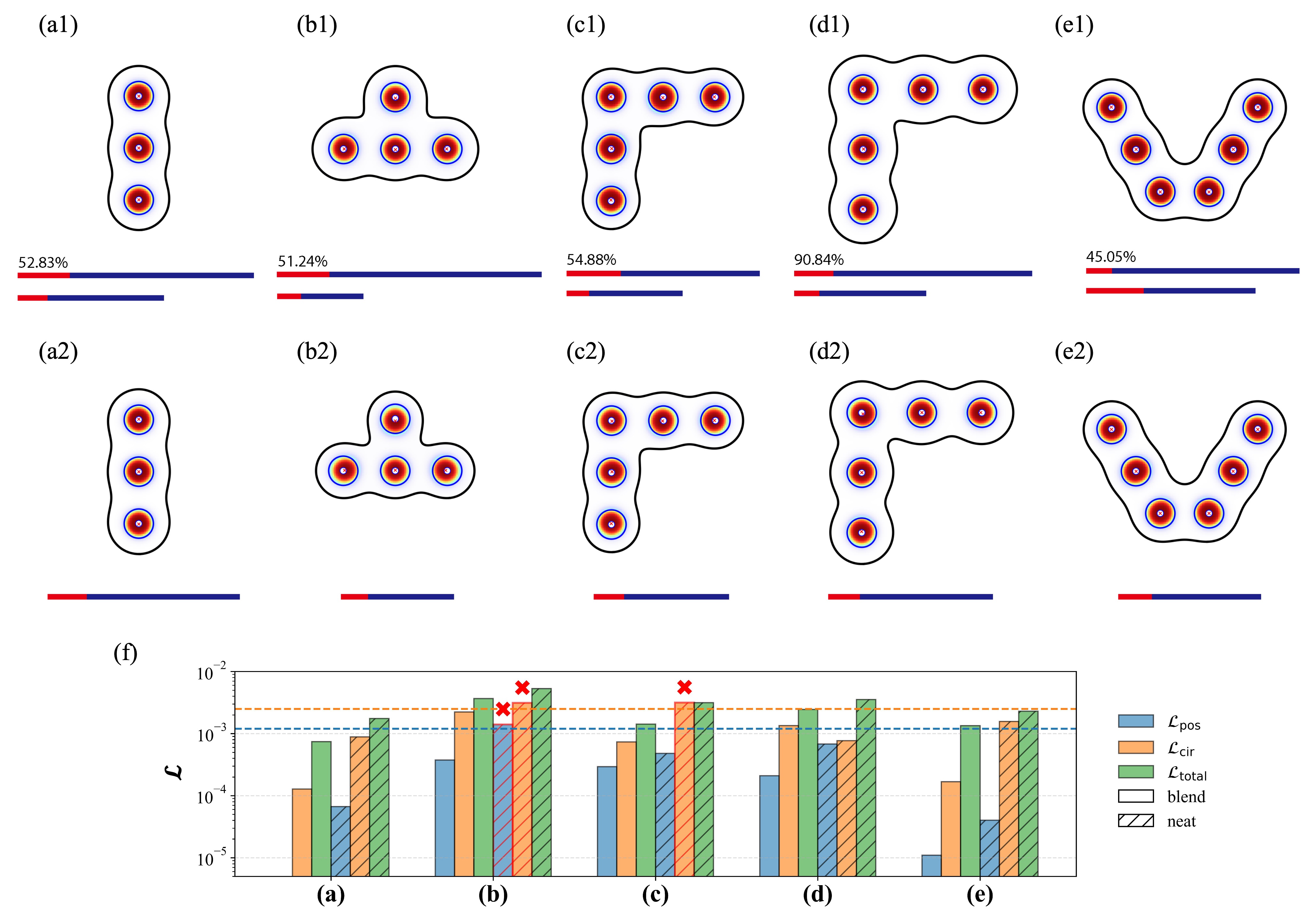}
    \caption{Self-assembled morphologies of the  
    optimized AB/AB blend (a1-e1) or neat AB diblock (a2-e2)
    in the optimized template 
    under curvature constraint: (a) triple-hole, (b) quadruple-hole, 
    (c,d) quintuple-hole, and (e) sextuple-hole. 
    (f) comparison of the objective functions for different multi-hole patterns.}
    \label{soft_co_opt}
\end{figure}

We performed the co-optimization of the template shape and the blending 
parameters for five multi-hole patterns, which correspond to patterns b, e, f, g and h in Figure \ref{co_opt}, respectively, 
after imposing the constraint to the variation of curvature 
of the template by setting $\tau \ge 0.35$ for Figures \ref{co_opt}b, f, g, and h,
while $\tau \ge 0.3$ for Figure \ref{co_opt}e.
For the blend, there are five parameters to optimize, including
the numbers of A-segments ($N_{\rm A_1}=f_{\rm A_1}N$ and $N_{\rm A_2}=f_{\rm A_2}N$),
the numbers of B-segments ($N_{\rm B_1}=f_{\rm B_1}N$ and $N_{\rm B_2}=f_{\rm B_2}N$)
and the concentration of one copolymer (\textit{i.e.} $\phi_1$). 
The typical optimized results
are presented in Figure \ref{soft_co_opt}a1-e1, 
where $\mathcal{L}_{\rm pos}=0$, $3.7\times 10^{-4}$, $2.9\times 10^{-4}$, $2.1\times 10^{-4}$, $1.1\times 10^{-5}$, and $\mathcal{L}_{\rm cir}=1.3\times 10^{-4}$, $2.2\times 10^{-3}$, $7.3\times 10^{-4}$, $1.3\times 10^{-3}$, $1.7\times 10^{-4}$, respectively.
All these optimal solutions exhibit reasonably high precisions, like $\mathcal{L}_{\rm pos}<1.2\times 10^{-3}$ and $\mathcal{L}_{\rm cir}<2.5\times 10^{-3}$.
More critically, all of these optimized templates show
noticeably milder variation of curvature 
($\tau=0.35$, $0.32$, $0.36$, $0.37$ and $0.36$) 
than their counterparts
in Figures \ref{co_opt}a, e, g, h and i achieved by the optimization process without
the constraint on the template shape 
($\tau=0.013$, $0.10$, $0.11$, $0.17$ and $0.15$). 
These results verify that the co-optimization strategy coupled with the shape
constraint is more applicable.

Additionally, when using neat AB diblock for co-optimization,
under the same constraints on the variation of curvature of the template, 
the optimization struggles to find satisfactory solutions with high precision
comparable to those of AB/AB blend for some multi-hole patterns.
As shown in Figure \ref{soft_co_opt}f, for the patterns
in Figures \ref{soft_co_opt}a2-e2 except for that in Figure \ref{soft_co_opt}d2, 
the optimal solutions obtained with neat AB diblock are overall inferior to those obtained with the AB/AB blend in terms of both $\mathcal{L}_{\rm pos}$ and $\mathcal{L}_{\rm cir}$. 
In fact, for patterns b2 and c2, the optimal solutions already exceed the acceptable tolerance range ($\mathcal{L}_{\rm pos}< 1.2\times 10^{-3}$ and $\mathcal{L}_{\rm cir}< 2.5\times 10^{-3}$). 
These results confirm that the AB/AB blend has better adaptability than the neat
AB diblock to template shape.

\section{Conclusions}

In summary, we have carried theoretical research on parameter optimization
for the directed self-assembly (DSA) of block copolymers, focusing on
the fabrication of various multi-hole patterns with precise size and
positions. To reduce the parameters of the guiding template
that need to be optimized, a Gaussian descriptor is proposed, which
only requires two parameters for describing each template.
Moreover, a multi-objective function is properly devised to
quantify the discrepancies between the self-assembled morphologies
and the target multi-hole patterns. Bayesian optimization, a gradient-free
black-box optimization method, is applied to optimize the
parameters for the minimal multi-objective function.
We first optimize the template shape (\textit{i.e.}, the two characteristic parameters of the Gaussian descriptor) for a pure cylinder-forming diblock copolymer melt to target
a specific double-hole pattern. It is demonstrated that the optimized template can guide the block copolymer to form the self-assembled morphology that precisely matches
the target double-hole pattern. These results testify that the Gaussian
descriptor is effective for describing the template shape, and the
multi-objective function is well devised for measuring the discrepancies
between the self-assembled morphology and the target multi-hole pattern.

Due to the limited adaptability of neat block copolymers to the template, the optimized templates may exhibit large variation of curvature, which is hard to be
manufactured. Based on this observation, we propose replacing the neat diblock copolymer with a binary blend composed of 
two different AB/AB diblock copolymers. 
By using the AB/AB blend with different B-blocks, the optimized template 
exhibits lower curvature variation, demonstrating a 
stronger adaptability of the blend to the templates shape than a neat copolymer system. 
Alternatively, we can also optimize the blend system to adapt to  
a specific template with reasonably lower curvature variation.
The results show that the binary blend exhibits a 
wide adjustable range of parameters, indicating high experimental feasibility. 
However, the binary blend cannot accommodate all templates. 
Therefore it is necessary to optimize both the template and the binary blend.

We perform the co-optimization of the template 
and the AB/AB blend for rather complex quintuple-hole pattern as a rigorous test case,
giving rise to a reasonably high precision.  
However, the optimization process sometimes sacrifices
the manufacturability of the template for persuiting high precision. 
Therefore, we further propose to impose a constraint on the variation of curvature
of the template during the optimization process by simply setting the
threshold of the characteristic parameter $\tau$.
Our results verify that under the constraint on the template shape
the cooptimization strategy can obtain the template shape with
reasonably good manufacturability that can lead to the self-assembly
morphology with high precision.

In large-scale sub-7nm technology node DSA-lithography, the adjacent VIAs 
on the layout will be grouped together and fabricated by the
self-assembly of block copolymers under the guide of a single template, followed by pattern transfer.
Although different templates may be used by different groups of multi-hole
patterns, a single blending formulation of block copolymers is
preferred for all groups of patterns
since the entire layout is uniformly coated through spin coating.
In principle, our co-optimization strategy can be directly applied
to simultaneously optimize the block copolymer blend and all the
templates for different groups of multi-hole patterns in the whole VIA layout.
Additionally, to enhance the adaptability of the blend to various templates further, we could consider slightly more complex blends, such as ternary blends or blends
containing ABA triblock copolymer.
Our work demonstrates that an efficient co-optimization strategy can
enhance the feasibility of DSA technique, thereby promoting its application.

\section{Methods}
\label{sec:method}

\subsection{Template descriptor}

How to parameterize the template shape is very
critical for its optimization, especially the number of parameters required
for describing the template shape. For simplicity, we assume that the template
is translationally invariant along the perpendicular direction, and accordingly
we just need to consider a two-dimensional shape. In mathematics, the shape
of a two-dimensional template is described by a closed curve. One usual way
of optimizing a closed curve is to divide it into many pieces and then optimize
the position of each pieces. This method involves dozens of tunable parameters, along with time-consuming self-consistent field theory (SCFT) calculations, making the entire optimization process computationally expensive.
Therefore, it is necessary to develop a model with as few parameters as possible, which can accurately describe the shape of DSA templates - particularly their curvature variation.

Since the local curvature of DSA templates for multi-hole patterns needs to conform to each nanohole, these templates share a common characteristic.
Specifically, the template shape approximates a circle near individual holes, while
its part between two adjacent holes is a smooth connection of two circular arcs.
For example, the typical DSA template for a double-hole pattern is a peanut-like shape,
which can be seen as two smoothly joined circles, whose curvature variation
increases with enlarging the separation of the two holes.
Based on the common characteristic, we propose a Gaussian distribution descriptor for DSA template shapes. The mathematical formulation of the multivariate Gaussian distribution is given below:
\begin{eqnarray}
f(\mathbf{x})=\sum_{i=1}^{n_{\rm hole}}\frac{1}{\sqrt{(2\pi)^k\det{\Sigma_{i}}}}\exp{\left(-\frac{1}{2}(\mathbf{x}-\mu_{i})^T\Sigma_{i}^{-1}(\mathbf{x}-\mu_{i})\right)}
\end{eqnarray}
where $\mathbf{x}$ denotes the coordinate vector in k-dimensional space, $n_{\rm hole}$ represents the number of target holes, and $\mu_i$, $\Sigma_i$ correspond to the mean vector and covariance matrix of the i-th Gaussian peak respectively.

\begin{figure}[htbp]
    \centering
    \includegraphics[width=1.0\textwidth]{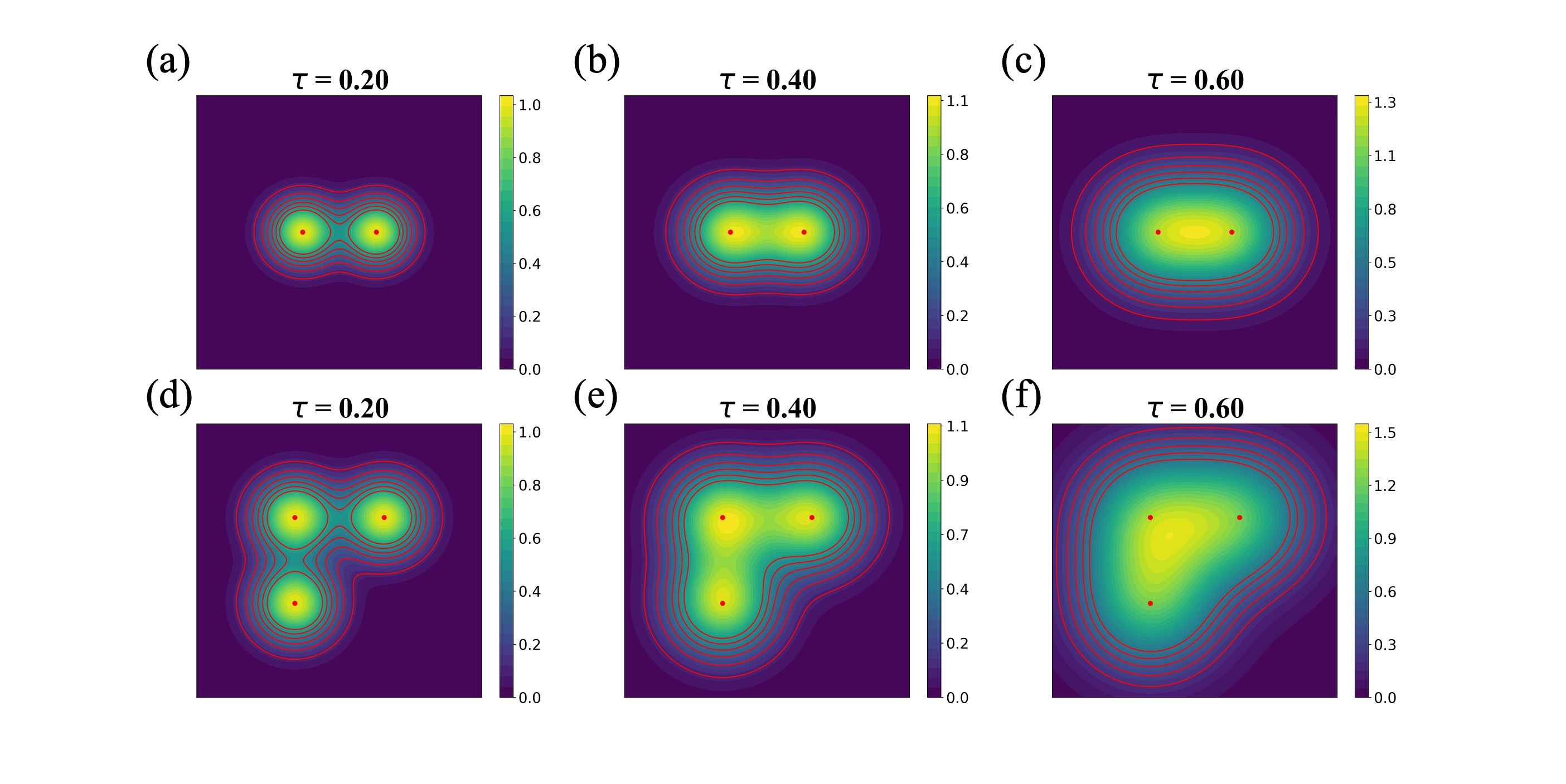}
    \caption{Superimposed distributions and contour profiles formed by 2D Gaussian peaks positioned at double-hole (a-c) and triple-hole (d-f) centers. For double-hole: Gaussian superimposed density distributions with thresholds of $\tau=0.2$, $0.4$, and $0.6$, respectively, where some typical contour profiles are shown with
    red curves.}
    \label{fig1}
\end{figure}

As illustrated in Figure \ref{fig1}, when targeting double-hole or triple-hole configurations, the placement of Gaussian peaks with varying variances (illustrated as $\tau$ temporarily and will be introduced later) at these target positions yields smooth and continuously varying contour profiles, which can represent different
template shapes. In other words, these contours naturally conform to the VIA holes layout geometry, satisfying the desired geometric characteristics for DSA guiding templates. Remarkably, only two parameters are required to generate these smooth DSA templates: the variance ($\sigma^2$) and the selected contour value ($\nu$).

However, in the original 2D Gaussian superposition distribution, the contour values vary globally with changes in variance, leading to parameter space shifts across different target multi-hole layout. This makes it difficult to establish a fixed parameter optimization range for arbitrary targets. To address this, we introduce a standardized scaling approach by employing a reference contour threshold parameter $\tau$. This parameter establishes a relationship with the scaling coefficient of Gaussian distribution, denoted as $A$, ensuring that the value ranges of superimposed Gaussian distributions under different configurations of target holes are approximately normalized to [0, 2], thereby simplifying the optimization process. The proof is as follows:

Under the assumption of ignoring covariance between $x$ and $y$ dimensions while maintaining equal variances ($\sigma^2$), the covariance matrix is reduced to:
\begin{eqnarray}
\Sigma=\ \left[\begin{matrix}\sigma^2&0\\0&\sigma^2\\\end{matrix}\right]
\end{eqnarray}
The probability density function of a single 2D Gaussian peak, scaled by coefficient $A$, is given by:
\begin{eqnarray}
F(x,y)=Af(x,y)=\frac{A}{2\pi\sigma^2}\exp{\left(-\frac{1}{2}\frac{(x-\mu_1)^2+(y-\mu_2)^2}{\sigma^2}\right)}
\end{eqnarray}
Assuming the contour at radius $R$ around the center is taken as the reference profile, its contour value is given by:
\begin{eqnarray}
{F(x,\ y)|}_{(x-\mu_1)^2+(y-\mu_2)^2=R^2}=\frac{A}{2\pi\sigma^2}\exp{\left(-\frac{R^2}{2\sigma^2}\right)}
\end{eqnarray}

We impose two key normalization conditions:

1. Peak normalization: The maximum probability density in the VIA center $(\mu_1, \mu_2)$ is fixed as 1.
\begin{eqnarray}
F(\mu_1,\ \mu_2)=\frac{A}{2\pi\sigma^2}=1
\end{eqnarray}

2. Contour threshold: The reference contour value at radius $R$ is fixed as $\tau$:
\begin{eqnarray}
{F(x,\ y)|}_{(x-\mu_1)^2+(y-\mu_2)^2=R^2}=\frac{A}{2\pi\sigma^2}\exp{\left(-\frac{R^2}{2\sigma^2}\right)}=\tau
\end{eqnarray}

Accordingly, we can derive the relationships between $\tau$ and $A, \sigma^2$:

\begin{eqnarray}
\sigma^2=-\frac{R^2}{2\ln{\tau}}\label{eq:sigma}
\end{eqnarray}
\begin{eqnarray}
A=-\frac{\pi R^2}{\ln{\tau}}\label{eq:A}
\end{eqnarray}

The reference radius $R$ can be fixed to a reasonable value, which can
be estimated from the DSA of pure AB diblock copolymer with the volume
fraction of A-block $f_{\rm{A}}=0.3$ for the fabrication
a single-hole pattern. Under strong-segregation approximation,
the radius of the target hole $R_{\rm{target}}$ and $R$ satisfies
the relationship $\pi R_{\rm{target}}^2 / \pi R^2 = f_{A}$,
giving rise to $R=R_{\rm{target}} /\sqrt{f_{\rm{A}}}$.
According to eq \ref{eq:sigma} and eq \ref{eq:A}, the overall contour curvature
can be controlled by $\tau$.
Moreover, for all complex target layouts, the resulting Gaussian probability density functions can be scaled to be within the range of $0<\nu<2.0$, as shown in Figure \ref{fig1}.

The contour can be determined using the Marching Squares algorithm based on the contour level $\nu$\cite{RN42}. Outside the contour that represents the template boundary,
the density distribution of wall is set to 1. For the implementation of
the masking method of SCFT, an appropriate wall density distribution is generated
so that the wall density inside the contour declines smoothly to 0. 
The Gaussian descriptor can not only significantly reduce the parametric dimensionality of the templates, but also be readily integrated with SCFT calculations,
thereby facilitating the co-optimization of the template shape and the block
copolymer system.

\subsection{Objective function}
To process the optimization, proper objective function is required to quantify the discrepancy between the self-assembled pattern and
the target pattern, which exists in terms of shape, size, position \textit{etc}.
Accordingly, we design a multi-objective function incorporating three metrics: circularity, center position accuracy, and redundancy distribution.

1. Circularity

The metric of circularity quantifies the deviation between the self-assembled
cylinders and the target circular holes. For multi-hole patterns, the cylindrical
domains often deviate from a circular shape when the template shape is inappropriate.
We simply use ellipses to fit these domain shapes by a six-point fitting algorithm, \cite{RN43}
followed by non-maximum suppression (NMS) to eliminate duplicate ellipses detected at identical locations due to algorithmic redundancy. 
Subsequently, we establish correspondence between all target holes and identified
ellipses, which is a classic assignment problem. By computing a distance matrix between all ellipse centers and target hole centers, we employ the Hungarian algorithm to minimize the total distance\cite{RN44}, thus achieving optimal matching between the target holes and the domains. The circularity objective function is formulated as:
\begin{eqnarray}
\mathcal{L}_{\rm{cir}}=\frac{1}{n_{\rm hole}}\sum_{i=1}^{n_{\rm hole}}\frac{(r_{i}-a_{i})^2+(r_{i}-b_{i})^2}{2r_{i}^2}
\end{eqnarray}
where $n_{\rm hole}$ denotes the number of target holes, $r_i$ represents the radius of the i-th target hole, while $a_i$ and $b_i$ indicate the semi-major and semi-minor axes of the matched ellipse, respectively. If insufficient ellipses are assigned to a target hole, default values of $a_i = b_i = 0$ are applied, resulting in a penalty score of 1.

Although domain shapes sometimes deviate significantly from ellipses, we find
that they all approach near-perfect circularity after the optimization process
is fully proceeded. In other words, if a domain substantially deviates from
an ellipse (where a circle is a special case of an ellipse), it must deviate significantly from the target circle as well, leading to a high penalty score for the deviation. The high penalty score is fed back into the optimization process to adjust the template shape accordingly. In a word, when the shape deviation is severe, the resulting penalty score need only be sufficiently high,
while their specific values have trivial impact on the optimization process.

2. Center position accuracy

The positional metric is introduced to quantify the deviation between the centers of all identified ellipses in the self-assembled morphology and their corresponding target holes. Inspired by the normalized Distance-IoU (DIoU) score from computer vision object detection tasks,\cite{RN45} we design the positional metric as follows:
\begin{eqnarray}
\mathcal{L}_{\rm{pos}}=\frac{1}{n_{\rm hole}}\sum_{i=1}^{n_{\rm hole}}\frac{(x_{i}-\widehat{x_{i}})^2+(y_{i}-\widehat{y_{i}})^2}{(x_{i, \rm{left}}-x_{i, \rm{right}})^2+(y_{i, \rm{up}}-y_{i, \rm{down}})^2}
\end{eqnarray}
where $(x_i,y_i)$ indicates the center coordinates of the i-th target hole, $(\widehat{x_i},\widehat{y_i})$ represents the center coordinates of the matched ellipse, and $(x_{i, \rm{left}},y_{i, \rm{up}})$, $(x_{i, \rm{right}},y_{i, \rm{down}})$ indicate the coordinates of two diagonal vertices of the minimum bounding rectangle that fully encloses the target hole and its corresponding ellipse. The closer the two centers are, the closer the positional metric approaches 0, while the denominator tends toward a constant. Conversely, the farther the ellipse is from the target circle, the larger the fraction becomes. When the distance between the centers approaches the length of the rectangle's diagonal, the fraction approaches 1.

3. Redundancy distribution

When the template is unreasonable, irregular redundant A-block domains beyond those
cylinders corresponding to the target holes may appear. These domains are not expected, but are difficult to describe using specific shapes. Therefore, they can be quantified using the Mean Squared Error (MSE) based on the difference in density distribution.
\begin{eqnarray}
\mathcal{L}_{\rm{rd}}=\frac{\int_{\Omega}{\max{\left(\rho_{\rm{actual}}\left(x\right)-\rho_{\rm{target}}\left(x\right),0\right)}dx}}{\int_{\Omega}{\rho_{\rm{actual}}\left(x\right)dx}}
\end{eqnarray}
In the above equation, $\Omega$ denotes the inside area of the template contour. $\rho_{\rm{actual}}(x)$ and $\rho_{\rm{target}}(x)$ represent the actual A-component density and the target density at position $x$, respectively. The target density is generated based on a $\tanh$ function centered at the desired position.
\begin{eqnarray}
\rho_{\rm{target}}(x)=\sum_{i=0}^{n_{\rm hole}}\frac{\tanh{(\alpha(r_i-|x-{x}_i|))}+1}{2}
\end{eqnarray}
where $x$ represents spatial location, while ${x}_i$ and $r_i$ denote the center position and radius of the i-th target hole, respectively. $\alpha$ is a heuristic parameter that determines the gradient of the distribution; its value can be empirically determined by $\chi N$. The above calculation only considers the redundant regions. The larger the proportion of redundant regions in the actual pattern, the closer the score will be to 1.

Finally, we obtain the multi-objective function by summing
the three metrics with appropriate weights:
\begin{eqnarray}
\mathcal{L}_{\rm{total}}=\alpha\mathcal{L}_{\rm{pos}}+\beta\mathcal{L}_{\rm{cir}}+\gamma\mathcal{L}_{\rm{rd}}
\end{eqnarray}
where $\alpha+\beta+\gamma=1$. With this multi-objective function,
we can establish optimization methods (such as Bayesian optimization, BO) to efficiently optimize both the block copolymer system and the template shape.

\subsection{Bayesian Optimization}
Bayesian optimization is a gradient-free global optimization algorithm that serves as a powerful solution for adaptive design problems. It is particularly useful when dealing with non-convex objective functions that are expensive to evaluate and have inaccessible derivatives, due to its outstanding ability to leverage the complete information from past optimization history. \cite{RN72, RN73}

The Bayesian optimization algorithm employs a surrogate model to approximate the true objective function, and uses an acquisition function to balance exploration and exploitation. This allows it to actively select evaluation points with the highest ``potential'', thereby avoiding unnecessary sampling.

In this work, we utilize the Bayesian optimization algorithm to efficiently search
the parameter space of both the template shape and the AB/AB blend, $\mathbf{X}$, for the parameter point $\mathbf{X}^*$ that corresponds to the global minimum of the total objective function $\mathcal{L}_{\rm{total}}$.
\begin{eqnarray}
\mathbf{X}^\ast=\arg{\min_{\mathbf{X}}{\mathcal{L}_{\rm{total}}(\mathbf{X})}}
\end{eqnarray}
where $\mathbf{X}$ denotes the parameters including template
parameters $\tau$ and $\nu$, as well as those of the AB/AB blend, including each block length,$f_{\rm A_1}N, f_{\rm A_2}N, f_{\rm B_1}N, f_{\rm B_2}N$ and volume fraction of first chain, $\phi_1$. To enhance the adaptability of the mixture to the template
dimension, $f_{\rm A_1}+f_{\rm B_1}=1$ and $f_{\rm A_2}+f_{\rm B_2}=1$ are not necessary.
In practice, we found that under certain extreme values of $\tau$, some corresponding values of $\nu$ could not generate viable template contours. These parameter combinations constitute an infeasible region. Therefore, we adopt a constrained Bayesian optimization algorithm,\cite{RN46} introducing a constraint function $c(\mathbf{X})$. A parameter set $\mathbf{X}$ is considered feasible only when $c(\mathbf{X})\le\lambda$. As a result, the optimization process is reformulated as follows:
\begin{eqnarray}
\mathbf{X}^\ast=\arg{\min_{\mathbf{X},c(\mathbf{X})\le\lambda}{\mathcal{L}_{\rm{total}}(\mathbf{X})}}.
\end{eqnarray}

Both $c(\mathbf{X})$ and $\mathcal{L}_{\rm{total}}(\mathbf{X})$ are modeled using Gaussian processes. In each iteration of the Bayesian optimization process, not only is $\mathcal{L}_{\rm{total}}(\mathbf{X})$ evaluated and the pair $(\mathbf{X},\mathcal{L}_{\rm{total}}(\mathbf{X}))$ is added to the observed set $T_l$, but $c(\mathbf{X})$ is also evaluated and the pair $(\mathbf{X},c(\mathbf{X}))$ is added to the observed set $T_c$, both of which are used to update the posterior of the Gaussian processes. Although infeasible points are not considered as candidates for the optimal solution, they are still included in $T_l$ and $T_c$ to improve the posterior estimates. These infeasible samples contribute to identifying the shape of $c(\mathbf{x})$, allowing the Gaussian process to better distinguish which regions of the parameter space are more likely to be feasible.

\begin{acknowledgement}

This work was supported by the Ministry of Science and Technology of
China (2024YFA1209000) and
the National Natural Science Foundation of China 
(Grant Nos. 52394272, 22333002, 22203018). 

\end{acknowledgement}

\begin{suppinfo}



\begin{itemize}
    \item Details on the Self-Consistent Field Theory modeling process can be found in S1.
\end{itemize}

\end{suppinfo}


\bibliography{reference}

\providecommand{\latin}[1]{#1}
\makeatletter
\providecommand{\doi}
  {\begingroup\let\do\@makeother\dospecials
  \catcode`\{=1 \catcode`\}=2 \doi@aux}
\providecommand{\doi@aux}[1]{\endgroup\texttt{#1}}
\makeatother
\providecommand*\mcitethebibliography{\thebibliography}
\csname @ifundefined\endcsname{endmcitethebibliography}  {\let\endmcitethebibliography\endthebibliography}{}
\begin{mcitethebibliography}{56}
\providecommand*\natexlab[1]{#1}
\providecommand*\mciteSetBstSublistMode[1]{}
\providecommand*\mciteSetBstMaxWidthForm[2]{}
\providecommand*\mciteBstWouldAddEndPuncttrue
  {\def\EndOfBibitem{\unskip.}}
\providecommand*\mciteBstWouldAddEndPunctfalse
  {\let\EndOfBibitem\relax}
\providecommand*\mciteSetBstMidEndSepPunct[3]{}
\providecommand*\mciteSetBstSublistLabelBeginEnd[3]{}
\providecommand*\EndOfBibitem{}
\mciteSetBstSublistMode{f}
\mciteSetBstMaxWidthForm{subitem}{(\alph{mcitesubitemcount})}
\mciteSetBstSublistLabelBeginEnd
  {\mcitemaxwidthsubitemform\space}
  {\relax}
  {\relax}

\bibitem[Wei \latin{et~al.}(2022)Wei, Han, Zhong, Xiao, Liu, and Xiang]{RN30}
Wei,~T.; Han,~Z.; Zhong,~X.; Xiao,~Q.; Liu,~T.; Xiang,~D. Two dimensional semiconducting materials for ultimately scaled transistors. \emph{IScience} \textbf{2022}, \emph{25}, 105160\relax
\mciteBstWouldAddEndPuncttrue
\mciteSetBstMidEndSepPunct{\mcitedefaultmidpunct}
{\mcitedefaultendpunct}{\mcitedefaultseppunct}\relax
\EndOfBibitem
\bibitem[Neisser(2021)]{RN67}
Neisser,~M. International roadmap for devices and systems lithography roadmap. \emph{J. Micro/Nanopatterning, Mater., Metrol.} \textbf{2021}, \emph{20}, 044601\relax
\mciteBstWouldAddEndPuncttrue
\mciteSetBstMidEndSepPunct{\mcitedefaultmidpunct}
{\mcitedefaultendpunct}{\mcitedefaultseppunct}\relax
\EndOfBibitem
\bibitem[Garner(2012)]{RN68}
Garner,~C.~M. Lithography for enabling advances in integrated circuits and devices. \emph{Philos. trans., Math. phys. eng. sci.} \textbf{2012}, \emph{370}, 4015--4041\relax
\mciteBstWouldAddEndPuncttrue
\mciteSetBstMidEndSepPunct{\mcitedefaultmidpunct}
{\mcitedefaultendpunct}{\mcitedefaultseppunct}\relax
\EndOfBibitem
\bibitem[Hummler \latin{et~al.}(2024)Hummler, Zhu, Behm, Matthes, He, Biabani, LaForge, Rollinger, Urone, and Kleemans]{RN70}
Hummler,~K.; Zhu,~Q.; Behm,~K.; Matthes,~L.; He,~Z.; Biabani,~O.; LaForge,~A.; Rollinger,~B.; Urone,~D.; Kleemans,~N. High-power EUV light sources (> 500w) for high throughput in next-generation EUV lithography tools. \emph{Proc. Opt. EUV Nanolithogr. XXXVII} \textbf{2024}, \emph{12953}, 309--316\relax
\mciteBstWouldAddEndPuncttrue
\mciteSetBstMidEndSepPunct{\mcitedefaultmidpunct}
{\mcitedefaultendpunct}{\mcitedefaultseppunct}\relax
\EndOfBibitem
\bibitem[Shintake and Naulleau(2025)Shintake, and Naulleau]{RN69}
Shintake,~T.; Naulleau,~P. Low-cost energy-efficient EUV lithography for advanced semiconductor manufacturing. \emph{Nat. Rev. Electr. Eng.} \textbf{2025}, \emph{2}, 2--3\relax
\mciteBstWouldAddEndPuncttrue
\mciteSetBstMidEndSepPunct{\mcitedefaultmidpunct}
{\mcitedefaultendpunct}{\mcitedefaultseppunct}\relax
\EndOfBibitem
\bibitem[Mohanty \latin{et~al.}(2023)Mohanty, Flores, and Chang]{RN52}
Mohanty,~S.; Flores,~E.; Chang,~C.-H. A study on EUV patterning with colloidal nanoparticles. \emph{Proc. Int. Conf. Extreme Ultraviolet Lithogr.} \textbf{2023}, \emph{12750}, 288--292\relax
\mciteBstWouldAddEndPuncttrue
\mciteSetBstMidEndSepPunct{\mcitedefaultmidpunct}
{\mcitedefaultendpunct}{\mcitedefaultseppunct}\relax
\EndOfBibitem
\bibitem[Thackeray(2011)]{RN50}
Thackeray,~J.~W. Materials challenges for sub-20-nm lithography. \emph{J. Micro/Nanolithogr., MEMS, MOEMS} \textbf{2011}, \emph{10}, 033009\relax
\mciteBstWouldAddEndPuncttrue
\mciteSetBstMidEndSepPunct{\mcitedefaultmidpunct}
{\mcitedefaultendpunct}{\mcitedefaultseppunct}\relax
\EndOfBibitem
\bibitem[Levinson(2025)]{RN51}
Levinson,~H.~J. Challenges and limits to patterning using extreme ultraviolet lithography. \emph{J. Micro/Nanopatterning, Mater., Metrol.} \textbf{2025}, \emph{24}, 011005--011005\relax
\mciteBstWouldAddEndPuncttrue
\mciteSetBstMidEndSepPunct{\mcitedefaultmidpunct}
{\mcitedefaultendpunct}{\mcitedefaultseppunct}\relax
\EndOfBibitem
\bibitem[Zhou \latin{et~al.}(2022)Zhou, Zhang, Li, Tang, Xiong, and Wang]{RN33}
Zhou,~H.; Zhang,~T.; Li,~S.; Tang,~M.; Xiong,~S.; Wang,~X. A via/contact layout decomposition method for directed self-assembly based on local optimization. \emph{Proc. Int. Workshop Adv. Pattern. Solut.} \textbf{2022}, \emph{1}, 1--4\relax
\mciteBstWouldAddEndPuncttrue
\mciteSetBstMidEndSepPunct{\mcitedefaultmidpunct}
{\mcitedefaultendpunct}{\mcitedefaultseppunct}\relax
\EndOfBibitem
\bibitem[Herr(2011)]{RN32}
Herr,~D.~J. Directed block copolymer self-assembly for nanoelectronics fabrication. \emph{J. Mater. Res.} \textbf{2011}, \emph{26}, 122--139\relax
\mciteBstWouldAddEndPuncttrue
\mciteSetBstMidEndSepPunct{\mcitedefaultmidpunct}
{\mcitedefaultendpunct}{\mcitedefaultseppunct}\relax
\EndOfBibitem
\bibitem[Cao \latin{et~al.}(2015)Cao, Zhang, Gu, Wang, and Lin]{cao2015designing}
Cao,~X.; Zhang,~L.; Gu,~J.; Wang,~L.; Lin,~J. Designing three-dimensional ordered structures from directed self-assembly of block copolymer films in topographical templates. \emph{Polymer} \textbf{2015}, \emph{72}, 10--20\relax
\mciteBstWouldAddEndPuncttrue
\mciteSetBstMidEndSepPunct{\mcitedefaultmidpunct}
{\mcitedefaultendpunct}{\mcitedefaultseppunct}\relax
\EndOfBibitem
\bibitem[Gu \latin{et~al.}(2021)Gu, Zhang, Zhang, and Lin]{gu2021epitaxial}
Gu,~J.; Zhang,~R.; Zhang,~L.; Lin,~J. Epitaxial assembly of nanoparticles in a diblock copolymer matrix: precise organization of individual nanoparticles into regular arrays. \emph{Macromolecules} \textbf{2021}, \emph{54}, 2561--2573\relax
\mciteBstWouldAddEndPuncttrue
\mciteSetBstMidEndSepPunct{\mcitedefaultmidpunct}
{\mcitedefaultendpunct}{\mcitedefaultseppunct}\relax
\EndOfBibitem
\bibitem[Kim \latin{et~al.}(2020)Kim, Jin, Yang, Han, Yun, Shin, Jeong, and Kim]{RN35}
Kim,~J.~H.; Jin,~H.~M.; Yang,~G.~G.; Han,~K.~H.; Yun,~T.; Shin,~J.~Y.; Jeong,~S.; Kim,~S.~O. Smart nanostructured materials based on self‐assembly of block copolymers. \emph{Adv. Funct. Mater.} \textbf{2020}, \emph{30}, 1902049\relax
\mciteBstWouldAddEndPuncttrue
\mciteSetBstMidEndSepPunct{\mcitedefaultmidpunct}
{\mcitedefaultendpunct}{\mcitedefaultseppunct}\relax
\EndOfBibitem
\bibitem[Pinto-Gómez \latin{et~al.}(2020)Pinto-Gómez, Pérez-Murano, Bausells, Villanueva, and Fernández-Regúlez]{RN54}
Pinto-Gómez,~C.; Pérez-Murano,~F.; Bausells,~J.; Villanueva,~L.~G.; Fernández-Regúlez,~M. Directed self-assembly of block copolymers for the fabrication of functional devices. \emph{Polymers} \textbf{2020}, \emph{12}, 2432\relax
\mciteBstWouldAddEndPuncttrue
\mciteSetBstMidEndSepPunct{\mcitedefaultmidpunct}
{\mcitedefaultendpunct}{\mcitedefaultseppunct}\relax
\EndOfBibitem
\bibitem[Xiong \latin{et~al.}(2016)Xiong, Wan, Ishida, Chapuis, Craig, Ruiz, and Nealey]{xiong2016directed}
Xiong,~S.; Wan,~L.; Ishida,~Y.; Chapuis,~Y.-A.; Craig,~G.~S.; Ruiz,~R.; Nealey,~P.~F. Directed self-assembly of triblock copolymer on chemical patterns for sub-10-nm nanofabrication via solvent annealing. \emph{ACS Nano} \textbf{2016}, \emph{10}, 7855--7865\relax
\mciteBstWouldAddEndPuncttrue
\mciteSetBstMidEndSepPunct{\mcitedefaultmidpunct}
{\mcitedefaultendpunct}{\mcitedefaultseppunct}\relax
\EndOfBibitem
\bibitem[Rasappa \latin{et~al.}(2018)Rasappa, Schulte, Ndoni, and Niemi]{RN53}
Rasappa,~S.; Schulte,~L.; Ndoni,~S.; Niemi,~T. Directed self-assembly of a high-chi block copolymer for the fabrication of optical nanoresonators. \emph{Nanoscale} \textbf{2018}, \emph{10}, 18306--18314\relax
\mciteBstWouldAddEndPuncttrue
\mciteSetBstMidEndSepPunct{\mcitedefaultmidpunct}
{\mcitedefaultendpunct}{\mcitedefaultseppunct}\relax
\EndOfBibitem
\bibitem[Mishra \latin{et~al.}(2022)Mishra, Lee, Kang, Kim, Choi, and Kim]{mishra2022gallol}
Mishra,~A.~K.; Lee,~J.; Kang,~S.; Kim,~E.; Choi,~C.; Kim,~J.~K. Gallol-based block copolymer with a high flory--huggins interaction parameter for next-generation lithography. \emph{Macromolecules} \textbf{2022}, \emph{55}, 10797--10803\relax
\mciteBstWouldAddEndPuncttrue
\mciteSetBstMidEndSepPunct{\mcitedefaultmidpunct}
{\mcitedefaultendpunct}{\mcitedefaultseppunct}\relax
\EndOfBibitem
\bibitem[Kwak \latin{et~al.}(2017)Kwak, Mishra, Lee, Lee, Choi, Maiti, Kim, and Kim]{kwak2017fabrication}
Kwak,~J.; Mishra,~A.~K.; Lee,~J.; Lee,~K.~S.; Choi,~C.; Maiti,~S.; Kim,~M.; Kim,~J.~K. Fabrication of sub-3 nm feature size based on block copolymer self-assembly for next-generation nanolithography. \emph{Macromolecules} \textbf{2017}, \emph{50}, 6813--6818\relax
\mciteBstWouldAddEndPuncttrue
\mciteSetBstMidEndSepPunct{\mcitedefaultmidpunct}
{\mcitedefaultendpunct}{\mcitedefaultseppunct}\relax
\EndOfBibitem
\bibitem[Wu \latin{et~al.}(2024)Wu, Luo, Dong, Li, Xu, Li, Li, Zhang, and Xiong]{RN15}
Wu,~Z.; Luo,~J.; Dong,~Q.; Li,~J.; Xu,~X.; Li,~Z.; Li,~W.; Zhang,~Y.; Xiong,~S. Quadruple-hole multiplication by directed self-assembly of block copolymer. \emph{Proc. Int. Workshop Adv. Pattern. Solut.} \textbf{2024}, \emph{13423}, 391--396\relax
\mciteBstWouldAddEndPuncttrue
\mciteSetBstMidEndSepPunct{\mcitedefaultmidpunct}
{\mcitedefaultendpunct}{\mcitedefaultseppunct}\relax
\EndOfBibitem
\bibitem[You \latin{et~al.}(2014)You, Park, Kim, Park, Seo, Lee, Jung, and Lee]{RN5}
You,~B.~K.; Park,~W.~I.; Kim,~J.~M.; Park,~K.-I.; Seo,~H.~K.; Lee,~J.~Y.; Jung,~Y.~S.; Lee,~K.~J. Reliable control of filament formation in resistive memories by self-assembled nanoinsulators derived from a block copolymer. \emph{ACS Nano} \textbf{2014}, \emph{8}, 9492--9502\relax
\mciteBstWouldAddEndPuncttrue
\mciteSetBstMidEndSepPunct{\mcitedefaultmidpunct}
{\mcitedefaultendpunct}{\mcitedefaultseppunct}\relax
\EndOfBibitem
\bibitem[Liu \latin{et~al.}(2018)Liu, Franke, Mignot, Xie, Yeung, Zhang, Chi, Zhang, Farrell, and Lai]{RN6}
Liu,~C.-C.; Franke,~E.; Mignot,~Y.; Xie,~R.; Yeung,~C.~W.; Zhang,~J.; Chi,~C.; Zhang,~C.; Farrell,~R.; Lai,~K. Directed self-assembly of block copolymers for 7 nanometre FinFET technology and beyond. \emph{Nat. Electron.} \textbf{2018}, \emph{1}, 562--569\relax
\mciteBstWouldAddEndPuncttrue
\mciteSetBstMidEndSepPunct{\mcitedefaultmidpunct}
{\mcitedefaultendpunct}{\mcitedefaultseppunct}\relax
\EndOfBibitem
\bibitem[Mokarian-Tabari \latin{et~al.}(2017)Mokarian-Tabari, Senthamaraikannan, Glynn, Collins, Cummins, Nugent, O’dwyer, and Morris]{RN7}
Mokarian-Tabari,~P.; Senthamaraikannan,~R.; Glynn,~C.; Collins,~T.~W.; Cummins,~C.; Nugent,~D.; O’dwyer,~C.; Morris,~M.~A. Large block copolymer self-assembly for fabrication of subwavelength nanostructures for applications in optics. \emph{Nano Lett.} \textbf{2017}, \emph{17}, 2973--2978\relax
\mciteBstWouldAddEndPuncttrue
\mciteSetBstMidEndSepPunct{\mcitedefaultmidpunct}
{\mcitedefaultendpunct}{\mcitedefaultseppunct}\relax
\EndOfBibitem
\bibitem[Suh \latin{et~al.}(2017)Suh, Kim, Moni, Xiong, Ocola, Zaluzec, Gleason, and Nealey]{RN71}
Suh,~H.~S.; Kim,~D.~H.; Moni,~P.; Xiong,~S.; Ocola,~L.~E.; Zaluzec,~N.~J.; Gleason,~K.~K.; Nealey,~P.~F. Sub-10-nm patterning via directed self-assembly of block copolymer films with a vapour-phase deposited topcoat. \emph{Nat. Nanotechnol.} \textbf{2017}, \emph{12}, 575--581\relax
\mciteBstWouldAddEndPuncttrue
\mciteSetBstMidEndSepPunct{\mcitedefaultmidpunct}
{\mcitedefaultendpunct}{\mcitedefaultseppunct}\relax
\EndOfBibitem
\bibitem[Verstraete \latin{et~al.}(2024)Verstraete, Suh, Van~Bel, Bak, Kim, Vallat, Bezard, Beggiato, and Beral]{RN47}
Verstraete,~L.; Suh,~H.~S.; Van~Bel,~J.; Bak,~B.-U.; Kim,~S.~E.; Vallat,~R.; Bezard,~P.; Beggiato,~M.; Beral,~C. Material and process optimization for EUV pattern rectification by DSA. \emph{Proc. Novel Pattern. Technol.} \textbf{2024}, \emph{12956}, 116--124\relax
\mciteBstWouldAddEndPuncttrue
\mciteSetBstMidEndSepPunct{\mcitedefaultmidpunct}
{\mcitedefaultendpunct}{\mcitedefaultseppunct}\relax
\EndOfBibitem
\bibitem[Kim \latin{et~al.}(2025)Kim, Kim, Kang, Her, Miyazaki, and Alperson]{RN48}
Kim,~D.-O.; Kim,~J.; Kang,~N.; Her,~Y.; Miyazaki,~S.; Alperson,~B. Improvement of EUV contact hole pattern rectification using novel directed self-assembly materials. \emph{Proc. Novel Pattern. Technol.} \textbf{2025}, \emph{13427}, 134270E\relax
\mciteBstWouldAddEndPuncttrue
\mciteSetBstMidEndSepPunct{\mcitedefaultmidpunct}
{\mcitedefaultendpunct}{\mcitedefaultseppunct}\relax
\EndOfBibitem
\bibitem[Van~Bel \latin{et~al.}(2024)Van~Bel, Verstraete, Suh, Bezard, Moussa, Santos, Her, and Gendt]{RN49}
Van~Bel,~J.; Verstraete,~L.; Suh,~H.~S.; Bezard,~P.; Moussa,~A.; Santos,~A.; Her,~Y.; Gendt,~S.~D. Rectification of extreme ultraviolet lithography patterns by directed self-assembly: a roughness and defectivity study. \emph{J. Micro/Nanopatterning, Mater., Metrol.} \textbf{2024}, \emph{23}, 043001\relax
\mciteBstWouldAddEndPuncttrue
\mciteSetBstMidEndSepPunct{\mcitedefaultmidpunct}
{\mcitedefaultendpunct}{\mcitedefaultseppunct}\relax
\EndOfBibitem
\bibitem[Bekaert \latin{et~al.}(2014)Bekaert, Doise, Kuppuswamy, Gronheid, Chan, Vandenberghe, Cao, and Her]{RN36}
Bekaert,~J.; Doise,~J.; Kuppuswamy,~V.-K.~M.; Gronheid,~R.; Chan,~B.~T.; Vandenberghe,~G.; Cao,~Y.; Her,~Y. Contact hole multiplication using grapho-epitaxy directed self-assembly: process choices, template optimization, and placement accuracy. \emph{Proc. 30th Eur. Mask Lithogr. Conf.} \textbf{2014}, \emph{9231}, 202--212\relax
\mciteBstWouldAddEndPuncttrue
\mciteSetBstMidEndSepPunct{\mcitedefaultmidpunct}
{\mcitedefaultendpunct}{\mcitedefaultseppunct}\relax
\EndOfBibitem
\bibitem[Cong \latin{et~al.}(2016)Cong, Zhang, Wang, and Lin]{cong2016understanding}
Cong,~Z.; Zhang,~L.; Wang,~L.; Lin,~J. Understanding the ordering mechanisms of self-assembled nanostructures of block copolymers during zone annealing. \emph{J. Chem. Phys.} \textbf{2016}, \emph{144}, 114901\relax
\mciteBstWouldAddEndPuncttrue
\mciteSetBstMidEndSepPunct{\mcitedefaultmidpunct}
{\mcitedefaultendpunct}{\mcitedefaultseppunct}\relax
\EndOfBibitem
\bibitem[Delachat \latin{et~al.}(2017)Delachat, Gharbi, Barros, Argoud, Lapeyre, Bos, Hazart, Pain, Monget, and Chevalier]{RN55}
Delachat,~F.; Gharbi,~A.; Barros,~P.~P.; Argoud,~M.; Lapeyre,~C.; Bos,~S.; Hazart,~J.; Pain,~L.; Monget,~C.; Chevalier,~X. Advanced surface affinity control for DSA contact hole shrink applications. \emph{Proc. Emerg. Pattern. Technol.} \textbf{2017}, \emph{10144}, 95--103\relax
\mciteBstWouldAddEndPuncttrue
\mciteSetBstMidEndSepPunct{\mcitedefaultmidpunct}
{\mcitedefaultendpunct}{\mcitedefaultseppunct}\relax
\EndOfBibitem
\bibitem[Wu \latin{et~al.}(2022)Wu, Dong, Xu, Li, Liu, Ji, Li, Manhua, and Xiong]{RN14}
Wu,~Z.; Dong,~Q.; Xu,~X.; Li,~Z.; Liu,~Y.; Ji,~S.; Li,~W.; Manhua,~S.; Xiong,~S. Improved processing window of contact hole with directed self-assembly of block copolymer blends. \emph{Proc. Int. Workshop Adv. Pattern. Solut.} \textbf{2022}, \emph{1}, 1--4\relax
\mciteBstWouldAddEndPuncttrue
\mciteSetBstMidEndSepPunct{\mcitedefaultmidpunct}
{\mcitedefaultendpunct}{\mcitedefaultseppunct}\relax
\EndOfBibitem
\bibitem[Shim and Shin(2016)Shim, and Shin]{RN9}
Shim,~S.; Shin,~Y. Mask optimization for directed self-assembly lithography: inverse DSA and inverse lithography. \emph{Proc. 21st Asia South Pac. Des. Autom. Conf.} \textbf{2016}, \emph{1}, 83--88\relax
\mciteBstWouldAddEndPuncttrue
\mciteSetBstMidEndSepPunct{\mcitedefaultmidpunct}
{\mcitedefaultendpunct}{\mcitedefaultseppunct}\relax
\EndOfBibitem
\bibitem[Wu \latin{et~al.}(2023)Wu, Dong, Luo, Yuan, Li, Liu, Ji, Li, Zhang, and Xiong]{RN8}
Wu,~Z.; Dong,~Q.; Luo,~J.; Yuan,~K.; Li,~Z.; Liu,~Y.; Ji,~S.; Li,~W.; Zhang,~Y.; Xiong,~S. Contact Hole Multiplication by Directed Self-Assembly of Block Copolymer with Homopolymer-Blending. \emph{Proc. Int. Workshop Adv. Pattern. Solut.} \textbf{2023}, \emph{1}, 1--4\relax
\mciteBstWouldAddEndPuncttrue
\mciteSetBstMidEndSepPunct{\mcitedefaultmidpunct}
{\mcitedefaultendpunct}{\mcitedefaultseppunct}\relax
\EndOfBibitem
\bibitem[Karageorgos \latin{et~al.}(2016)Karageorgos, Ryckaert, Gronheid, Tung, Wong, Karageorgos, Croes, Bekaert, Vandenberghe, and Stucchi]{RN13}
Karageorgos,~I.; Ryckaert,~J.; Gronheid,~R.; Tung,~M.~C.; Wong,~H.-S.~P.; Karageorgos,~E.; Croes,~K.; Bekaert,~J.; Vandenberghe,~G.; Stucchi,~M. Design method and algorithms for directed self-assembly aware via layout decomposition in sub-7 nm circuits. \emph{J. Micro/Nanolithogr., MEMS, MOEMS} \textbf{2016}, \emph{15}, 043506\relax
\mciteBstWouldAddEndPuncttrue
\mciteSetBstMidEndSepPunct{\mcitedefaultmidpunct}
{\mcitedefaultendpunct}{\mcitedefaultseppunct}\relax
\EndOfBibitem
\bibitem[Cheng \latin{et~al.}(2010)Cheng, Sanders, Truong, Harrer, Friz, Holmes, Colburn, and Hinsberg]{RN37}
Cheng,~J.~Y.; Sanders,~D.~P.; Truong,~H.~D.; Harrer,~S.; Friz,~A.; Holmes,~S.; Colburn,~M.; Hinsberg,~W.~D. Simple and versatile methods to integrate directed self-assembly with optical lithography using a polarity-switched photoresist. \emph{ACS Nano} \textbf{2010}, \emph{4}, 4815--4823\relax
\mciteBstWouldAddEndPuncttrue
\mciteSetBstMidEndSepPunct{\mcitedefaultmidpunct}
{\mcitedefaultendpunct}{\mcitedefaultseppunct}\relax
\EndOfBibitem
\bibitem[Zhang \latin{et~al.}(2019)Zhang, Zhang, Lin, and Lin]{zhang2019customizing}
Zhang,~R.; Zhang,~L.; Lin,~J.; Lin,~S. Customizing topographical templates for aperiodic nanostructures of block copolymers via inverse design. \emph{Phys. Chem. Chem. Phys.} \textbf{2019}, \emph{21}, 7781--7788\relax
\mciteBstWouldAddEndPuncttrue
\mciteSetBstMidEndSepPunct{\mcitedefaultmidpunct}
{\mcitedefaultendpunct}{\mcitedefaultseppunct}\relax
\EndOfBibitem
\bibitem[Matsen(2001)]{RN38}
Matsen,~M.~W. The standard Gaussian model for block copolymer melts. \emph{J. Phys.:Condens. Matter} \textbf{2001}, \emph{14}, R21\relax
\mciteBstWouldAddEndPuncttrue
\mciteSetBstMidEndSepPunct{\mcitedefaultmidpunct}
{\mcitedefaultendpunct}{\mcitedefaultseppunct}\relax
\EndOfBibitem
\bibitem[Yang \latin{et~al.}(2022)Yang, Dong, Liu, and Li]{RN39}
Yang,~J.; Dong,~Q.; Liu,~M.; Li,~W. Universality and specificity in the self-assembly of cylinder-forming block copolymers under cylindrical confinement. \emph{Macromolecules} \textbf{2022}, \emph{55}, 2171--2181\relax
\mciteBstWouldAddEndPuncttrue
\mciteSetBstMidEndSepPunct{\mcitedefaultmidpunct}
{\mcitedefaultendpunct}{\mcitedefaultseppunct}\relax
\EndOfBibitem
\bibitem[Iwama \latin{et~al.}(2015)Iwama, Laachi, Delaney, and Fredrickson]{RN77}
Iwama,~T.; Laachi,~N.; Delaney,~K.~T.; Fredrickson,~G.~H. Computational study of directed self-assembly for contact-hole shrink and multiplication. \emph{J. Micro/Nanolithogr., MEMS, MOEMS} \textbf{2015}, \emph{14}, 013501\relax
\mciteBstWouldAddEndPuncttrue
\mciteSetBstMidEndSepPunct{\mcitedefaultmidpunct}
{\mcitedefaultendpunct}{\mcitedefaultseppunct}\relax
\EndOfBibitem
\bibitem[Doise \latin{et~al.}(2019)Doise, Bezik, Hori, de~Pablo, and Gronheid]{RN41}
Doise,~J.; Bezik,~C.; Hori,~M.; de~Pablo,~J.~J.; Gronheid,~R. Influence of homopolymer addition in templated assembly of cylindrical block copolymers. \emph{ACS Nano} \textbf{2019}, \emph{13}, 4073--4082\relax
\mciteBstWouldAddEndPuncttrue
\mciteSetBstMidEndSepPunct{\mcitedefaultmidpunct}
{\mcitedefaultendpunct}{\mcitedefaultseppunct}\relax
\EndOfBibitem
\bibitem[Bosse \latin{et~al.}(2007)Bosse, Garcia-Cervera, and Fredrickson]{RN24}
Bosse,~A.~W.; Garcia-Cervera,~C.~J.; Fredrickson,~G.~H. Microdomain ordering in laterally confined block copolymer thin films. \emph{Macromolecules} \textbf{2007}, \emph{40}, 9570--9581\relax
\mciteBstWouldAddEndPuncttrue
\mciteSetBstMidEndSepPunct{\mcitedefaultmidpunct}
{\mcitedefaultendpunct}{\mcitedefaultseppunct}\relax
\EndOfBibitem
\bibitem[Matsen(1997)]{RN22}
Matsen,~M. Thin films of block copolymer. \emph{J. Chem. Phys.} \textbf{1997}, \emph{106}, 7781--7791\relax
\mciteBstWouldAddEndPuncttrue
\mciteSetBstMidEndSepPunct{\mcitedefaultmidpunct}
{\mcitedefaultendpunct}{\mcitedefaultseppunct}\relax
\EndOfBibitem
\bibitem[Xia and Li(2019)Xia, and Li]{RN25}
Xia,~Y.; Li,~W. Defect-free hexagonal patterns formed by AB diblock copolymers under triangular confinement. \emph{Polymer} \textbf{2019}, \emph{166}, 21--26\relax
\mciteBstWouldAddEndPuncttrue
\mciteSetBstMidEndSepPunct{\mcitedefaultmidpunct}
{\mcitedefaultendpunct}{\mcitedefaultseppunct}\relax
\EndOfBibitem
\bibitem[Zhang \latin{et~al.}(2022)Zhang, Yang, and Li]{RN26}
Zhang,~L.; Yang,~J.; Li,~W. Emergence of Multi-strand Helices from the Self-Assembly of AB-Type Multiblock Copolymer under Cylindrical Confinement. \emph{Macromolecules} \textbf{2022}, \emph{55}, 9334--9343\relax
\mciteBstWouldAddEndPuncttrue
\mciteSetBstMidEndSepPunct{\mcitedefaultmidpunct}
{\mcitedefaultendpunct}{\mcitedefaultseppunct}\relax
\EndOfBibitem
\bibitem[Latypov \latin{et~al.}(2014)Latypov, Garner, Preil, Schmid, Wang, Xu, and Zou]{RN40}
Latypov,~A.; Garner,~G.; Preil,~M.; Schmid,~G.; Wang,~W.-L.; Xu,~J.; Zou,~Y. Computational simulations and parametric studies for directed self-assembly process development and solution of the inverse directed self-assembly problem. \emph{Jpn. J. Appl. Phys.} \textbf{2014}, \emph{53}, 06JC01\relax
\mciteBstWouldAddEndPuncttrue
\mciteSetBstMidEndSepPunct{\mcitedefaultmidpunct}
{\mcitedefaultendpunct}{\mcitedefaultseppunct}\relax
\EndOfBibitem
\bibitem[Ouaknin \latin{et~al.}(2018)Ouaknin, Laachi, Delaney, Fredrickson, and Gibou]{RN11}
Ouaknin,~G.~Y.; Laachi,~N.; Delaney,~K.; Fredrickson,~G.~H.; Gibou,~F. Level-set strategy for inverse DSA-lithography. \emph{J. Comput. Phys.} \textbf{2018}, \emph{375}, 1159--1178\relax
\mciteBstWouldAddEndPuncttrue
\mciteSetBstMidEndSepPunct{\mcitedefaultmidpunct}
{\mcitedefaultendpunct}{\mcitedefaultseppunct}\relax
\EndOfBibitem
\bibitem[Bochkov and Gibou(2024)Bochkov, and Gibou]{RN12}
Bochkov,~D.; Gibou,~F. A Non-parametric Gradient-Based Shape Optimization Approach for Solving Inverse Problems in Directed Self-Assembly of Block Copolymers. \emph{Commun. Appl. Math. Comput.} \textbf{2024}, \emph{6}, 1472--1489\relax
\mciteBstWouldAddEndPuncttrue
\mciteSetBstMidEndSepPunct{\mcitedefaultmidpunct}
{\mcitedefaultendpunct}{\mcitedefaultseppunct}\relax
\EndOfBibitem
\bibitem[Wang \latin{et~al.}(2025)Wang, Li, Zeng, and Zhang]{RN56}
Wang,~H.; Li,~S.; Zeng,~J.; Zhang,~T. Accelerating polymer self-consistent field simulation and inverse DSA-lithography with deep neural networks. \emph{J. Chem. Phys.} \textbf{2025}, \emph{162}, 104105\relax
\mciteBstWouldAddEndPuncttrue
\mciteSetBstMidEndSepPunct{\mcitedefaultmidpunct}
{\mcitedefaultendpunct}{\mcitedefaultseppunct}\relax
\EndOfBibitem
\bibitem[Fühner \latin{et~al.}(2016)Fühner, Michalak, Welling, Orozco-Rey, Müller, and Erdmann]{RN75}
Fühner,~T.; Michalak,~P.; Welling,~U.; Orozco-Rey,~J.~C.; Müller,~M.; Erdmann,~A. An integrated source/mask/DSA optimization approach. \emph{Proc. Opt. Microlithogr. XXIX} \textbf{2016}, \emph{9780}, 65--76\relax
\mciteBstWouldAddEndPuncttrue
\mciteSetBstMidEndSepPunct{\mcitedefaultmidpunct}
{\mcitedefaultendpunct}{\mcitedefaultseppunct}\relax
\EndOfBibitem
\bibitem[Lorensen and Cline(1987)Lorensen, and Cline]{RN42}
Lorensen,~W.~E.; Cline,~H.~E. Marching cubes: A high resolution 3D surface construction algorithm. \emph{SIGGRAPH Comput. Graph.} \textbf{1987}, \emph{21}, 163–169\relax
\mciteBstWouldAddEndPuncttrue
\mciteSetBstMidEndSepPunct{\mcitedefaultmidpunct}
{\mcitedefaultendpunct}{\mcitedefaultseppunct}\relax
\EndOfBibitem
\bibitem[Fitzgibbon \latin{et~al.}(1999)Fitzgibbon, Pilu, and Fisher]{RN43}
Fitzgibbon,~A.; Pilu,~M.; Fisher,~R.~B. Direct least square fitting of ellipses. \emph{IEEE Trans. Pattern Anal. Mach. Intell.} \textbf{1999}, \emph{21}, 476--480\relax
\mciteBstWouldAddEndPuncttrue
\mciteSetBstMidEndSepPunct{\mcitedefaultmidpunct}
{\mcitedefaultendpunct}{\mcitedefaultseppunct}\relax
\EndOfBibitem
\bibitem[Kuhn(1955)]{RN44}
Kuhn,~H.~W. The Hungarian method for the assignment problem. \emph{Nav. Res. Logist. Q.} \textbf{1955}, \emph{2}, 83--97\relax
\mciteBstWouldAddEndPuncttrue
\mciteSetBstMidEndSepPunct{\mcitedefaultmidpunct}
{\mcitedefaultendpunct}{\mcitedefaultseppunct}\relax
\EndOfBibitem
\bibitem[Zheng \latin{et~al.}(2020)Zheng, Wang, Liu, Li, Ye, and Ren]{RN45}
Zheng,~Z.; Wang,~P.; Liu,~W.; Li,~J.; Ye,~R.; Ren,~D. Distance-IoU loss: Faster and better learning for bounding box regression. \emph{Proc. AAAI Conf. Artif. Intell.} \textbf{2020}, \emph{34}, 12993--13000\relax
\mciteBstWouldAddEndPuncttrue
\mciteSetBstMidEndSepPunct{\mcitedefaultmidpunct}
{\mcitedefaultendpunct}{\mcitedefaultseppunct}\relax
\EndOfBibitem
\bibitem[Dong \latin{et~al.}(2023)Dong, Gong, Yuan, Jiang, Zhang, and Li]{RN72}
Dong,~Q.; Gong,~X.; Yuan,~K.; Jiang,~Y.; Zhang,~L.; Li,~W. Inverse design of complex block copolymers for exotic self-assembled structures based on Bayesian optimization. \emph{ACS Macro Lett.} \textbf{2023}, \emph{12}, 401--407\relax
\mciteBstWouldAddEndPuncttrue
\mciteSetBstMidEndSepPunct{\mcitedefaultmidpunct}
{\mcitedefaultendpunct}{\mcitedefaultseppunct}\relax
\EndOfBibitem
\bibitem[Dong \latin{et~al.}(2024)Dong, Xu, Song, Qiang, Cao, and Li]{RN73}
Dong,~Q.; Xu,~Z.; Song,~Q.; Qiang,~Y.; Cao,~Y.; Li,~W. Automated search strategy for novel ordered structures of block copolymers. \emph{ACS Macro Lett.} \textbf{2024}, \emph{13}, 987--993\relax
\mciteBstWouldAddEndPuncttrue
\mciteSetBstMidEndSepPunct{\mcitedefaultmidpunct}
{\mcitedefaultendpunct}{\mcitedefaultseppunct}\relax
\EndOfBibitem
\bibitem[Gardner \latin{et~al.}(2014)Gardner, Kusner, Xu, Weinberger, and Cunningham]{RN46}
Gardner,~J.~R.; Kusner,~M.~J.; Xu,~Z.~E.; Weinberger,~K.~Q.; Cunningham,~J.~P. Bayesian optimization with inequality constraints. \emph{ICML} \textbf{2014}, \emph{32}, 937--945\relax
\mciteBstWouldAddEndPuncttrue
\mciteSetBstMidEndSepPunct{\mcitedefaultmidpunct}
{\mcitedefaultendpunct}{\mcitedefaultseppunct}\relax
\EndOfBibitem
\end{mcitethebibliography}

\end{document}